\documentclass[aps,prb,reprint,amsmath,amssymb]{revtex4-2} 
\usepackage[utf8]{inputenc}
\usepackage{braket,hyperref,graphicx}
\usepackage[dvipsnames]{xcolor}

\graphicspath{{./}{./figures/}}

\def\L{ {\mathrm{L}} }

\def\R{ {\mathrm{R}} }
\def\d{ {\mathrm{d}} }
\def\e{ {\mathrm{e}} }

\def\i{ {\mathrm{i}} }

\def\vk{ {\boldsymbol{k}} }

\def\vp{ {\boldsymbol{p}} }

\def\vr{ {\boldsymbol{r}} }

\def\vv{ {\boldsymbol{v}} }
\def\vw{ {\boldsymbol{w}} }

\newcommand{\bsg}{\boldsymbol{\sigma}}

\newcommand{\beq}{\begin{equation}}
\newcommand{\eeq}{\end{equation}}

\newcommand{\abs}[1]{\left| #1 \right|}

\newcommand{\nullv}{\mathbf{0}}

\newcommand{\diag}[1]{\text{diag}\left\{ #1 \right\}}

\newcommand{\cmplx}{\mathbb{C}}

\newcommand{\ve}{\varepsilon}

\newcommand{\dg}{{\dagger}}

\newcommand{\hlt}{\mathcal{H}}

\newcommand{\green}{\mathcal{G}}

\newcommand{\ie}{i.e.}

\newcommand\Tmat{\mathcal{T}}
\newcommand\Kmat{\mathcal{K}}

\newcommand{\ia}{\mathrm{intra}}
\newcommand{\ir}{\mathrm{inter}}

\newcommand\ktrans{\vk}

\newcommand{\td}{\tau_\mathrm{dwell}}
\newcommand{\tl}{\tau_\mathrm{life}}
\newcommand{\mfp}{l_\mathrm{mfp}}
\newcommand{\lir}{l_\mathrm{arc}}
\newcommand{\ba}{B_\mathrm{arc}}
\newcommand{\tu}{t}
\newcommand{\la}{L_\mathrm{arc}}

\begin{document}

\title{Magnetotransport across Weyl semimetal grain boundaries}

\author{Haoyang Tian}
\email{haoyang.tian@uni-koeln.de}
\author{Vatsal Dwivedi}
\author{Adam Yanis Chaou}
\author{Maxim Breitkreiz}
\email{breitkr@physik.fu-berlin.de}
\affiliation{Freie Universität Berlin, Dahlem Center for Complex Quantum Systems and Fachbereich Physik, Arnimallee 14, 14195 Berlin 
}

\begin{abstract}
A clean interface between two Weyl semimetals features a universal, field-linear tunnel magnetoconductance of $(e^2/h)N_\mathrm{ho}$ per magnetic flux quantum, where $N_\mathrm{ho}$ is the number of chirality-preserving topological interface Fermi arcs. 
In this work we show that the linearity of the magnetoconductance is robust with to interface disorder. The slope of the magnetoconductance changes at a characteristic field strength $B_\mathrm{arc}$ --- the field strength for which the time taken to traverse the Fermi arc due to the Lorentz force is equal to the mean inter-arc scattering time. For fields much larger than $B_\mathrm{arc}$, the magnetoconductance is unaffected by disorder. For fields much smaller than $B_\mathrm{arc}$, the slope is no longer determined by $N_\mathrm{ho}$ but by the simple fraction $N_\L N_\R/(N_\L+N_\R)$, where $N_\L$ and $N_\R$ are the numbers of Weyl-node pairs in the left and right Weyl semimetal, respectively. We also consider the effect of spatially correlated disorder potentials, where we find that $B_\mathrm{arc}$ decreases exponentially with increasing correlation length. 
Our results provide a possible explanation for the recently observed robustness of the negative linear magnetoresistance in grained Weyl semimetals. 
\end{abstract}

\maketitle

\section{Introduction}

Weyl semimetals (WSMs) have emerged as a paradigmatic platform for exploring gapless topological phases of matter. 
Characterized by topologically protected band crossings,
termed Weyl nodes \cite{Wan2011,Armitage2018}, the low-energy physics is governed by the Weyl Hamiltonian $H = \chi\,\vp\cdot\bsg$, where $\chi=\pm1$ denotes the chirality, $\vp$ the momentum, and $\bsg$ the vector of Pauli matrices corresponding to (pseudo) spin. Weyl nodes always occur in pairs of opposite chirality and can only be gapped out by pairwise annihilation. 

WSMs exhibit a range of anomalous transport phenomena rooted in their chiral electronic structure.
Among the most striking features are the negative longitudinal magnetoresistance induced by the chiral anomaly.
In the presence of an applied magnetic field, the chiral anomaly\cite{Adler1969,Bell1969} of Weyl nodes manifests in the lowest Landau level: The Landau quantization of the Weyl node leads to a chiral zeroth Landau level moving parallel or antiparallel to the magnetic field depending on the chirality. The resulting chiral magnetic effect is theoretically expected to manifest as a positive longitudinal magnetoconductance \cite{Son2013a, Burkov2017, Xiong2015,Altland2016}. 
Experimentally identifying the chiral magnetic effect remains challenging due to extrinsic effects, such as current jetting, the fact that Weyl nodes do not typically reside at the Fermi level, and are not the sole charge carriers in the system \cite{Reis2016, Naumann2020, Lv2021}.

\begin{figure}
	\includegraphics[width=0.8\columnwidth]{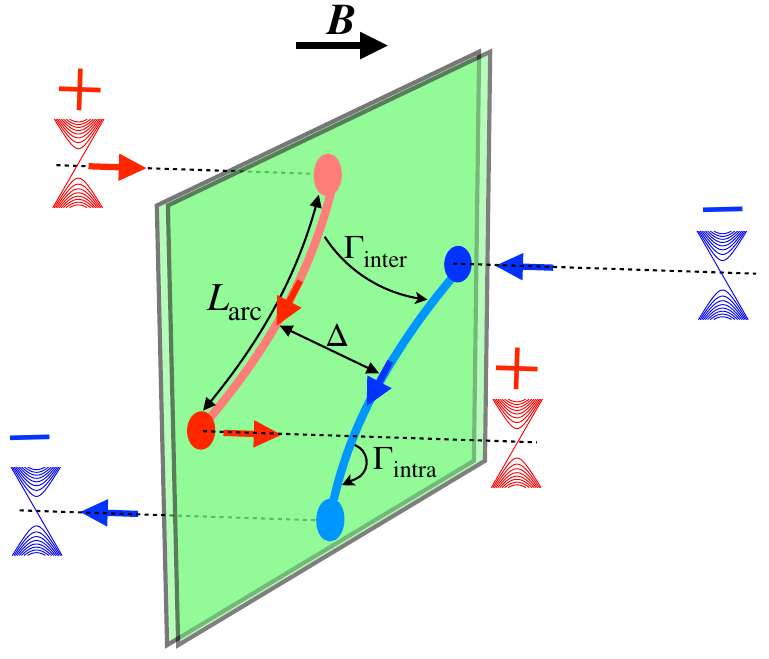}	
	\caption{WSM interface in momentum space (green area)  with two interface Fermi arcs (red/blue lines) connecting the projections of Weyl nodes (red/blue circles for positive/negative chirality). Intra- and inter-arc scattering is depicted, as well as the particle flow (red/blue arrows) in the presence of a magnetic field $\boldsymbol{B}$ normal to the interface.}
	\label{fig:1}    
\end{figure}

While most such experiments are performed on single crystals,  Ref.\ \cite{Zhang2023} reports the observation of a particularly robust chiral magnetic effect in a grained WSM with randomly oriented crystallites. On the theoretical side, some of us have recently shown that the chiral magnetic effect can occur at interfaces between different or differently oriented WSMs \cite{Chaou2023, Chaou2023a} in the form of a universal tunnel magnetoconductance that does not depend on the Fermi momentum and velocity of the Weyl fermions,
\begin{equation}
	G_0(B) =  \frac{e^3}{h^2} \abs{ \boldsymbol{A}\cdot \boldsymbol{B}},
\label{cond0}
\end{equation}
where $\abs{ \boldsymbol{A}\cdot \boldsymbol{B}}$ is the magnetic flux through the interface area $\boldsymbol{A}$.
While the original chiral magnetic effect is rooted only in the bulk Weyl fermions, the tunnel magnetoconductance across an interface involves topological interface states. These are related to the well-established WSM surface states, called Fermi arcs, which are open contours of chiral modes connecting the projections of \textit{opposite-chirality} bulk Weyl nodes on the surface Brillouin zone \cite{Wan2011}. Fermi arcs also exist at interfaces between WSMs, where they can connect Weyl nodes of the \textit{same chirality} on opposite sides of the interface \cite{Dwivedi2018,Murthy2020,Abdulla2021,Kaushik2022,Buccheri2022,Mathur2023,Kundu2023,Chaou2023}, as illustrated in Fig.\ \ref{fig:1}.
In the presence of a magnetic field perpendicular to the interface, such chirality-preserving (homochiral) interface Fermi arcs transmit the chiral Landau level of bulk Weyl fermions, while filtering out other modes, leading to the universal tunnel magnetoconductance $G(B)=N_\mathrm{ho}G_0(B)$, where $N_\mathrm{ho}$ is the number of pairs of homochiral Fermi-arc. This mechanism can always be extended to a setting consisting of a series of interfaces\cite{Breitkreiz2023}, thus leading to a chiral magnetic effect of a grained WSM. 

In this work we investigate how robust this Fermi-arc-mediated magnetotransport across interfaces is to interface disorder. While it is known that surface Fermi arcs contribute to in-plane transport in a disorder-robust fashion \cite{Breitkreiz2019, Piskunow2021,Lanzillo2024,Leahy2024,Khan2025}, the robustness of the Fermi-arc-mediated magnetotransport across a WSM interface is not obvious. To address this question, we develop a hybrid approach consisting of both Landauer-Bütticker and semiclassical Boltzmann transport formalisms. The former accounts for the chiral anomaly, while the latter incorporates the effects of disorder. We corroborate our analytical predictions by way of lattice simulations on a tight-binding model. Our main result is that the linearity of the magnetoconductance is robust and features two different slopes in the limits of fields being much larger and much smaller than $B_\mathrm{arc}$ --- a characteristic field strength at which the time it takes to traverse a Fermi arc driven by the Lorentz force is equal to the mean time between inter-arc scattering events. For fields much larger than $B_\mathrm{arc}$, the magnetoconductance is identical to the clean-interface case. For fields much smaller than $B_\mathrm{arc}$, the magnetoconductance becomes a simple fraction of $G_0(B)$.
The resulting linear, slope-changing positive magnetoconductance qualitatively agrees well with the experimental measurement of the magnetoconductance in a grained Weyl semimetal \cite{Zhang2023}.

This article is organized as follows. In Sec. \ref{sec:minimal_model} we introduce the Landauer-Boltzmann approach to the tunnel magnetoconductance behavior of disordered interfaces and discuss the generic properties of the conductance. In Sec. \ref{sec:numerics} we test and exemplify the transport behavior on a minimal lattice model introduced in Sec.\ 
\ref{sec:kwant}. Specifically,
in Sec. \ref{sec:LB_approach} we deploy our hybrid approach on the lattice model and compare with full-quantum transport simulations. In Sec. \ref{sec:mb} we explore the special case of weakly coupled interfaces with magnetic breakdown. In Sec.\ \ref{sec:ncl} we discuss the disorder-activated contribution of non-chiral Landau levels. We conclude in Sec.\ \ref{sec:conc}.

\section{Analytical model}
\label{sec:minimal_model}

We start by considering a minimal model of  an interface between two WSMs, each featuring a single pair of opposite-chirality Weyl nodes whose position changes across the interface, as sketched in Fig.\ \ref{fig:1}. Let both Fermi arcs be of length $\la$ and consider an applied magnetic field $B$ perpendicular to the interface. In the bulk, the magnetic field gives rise to Landau levels at the Weyl nodes with a chiral lowest Landau level that moves parallel/antiparallel to the field depending on the Weyl-node chirality. 
Bulk currents from the chiral Landau levels on either side of the interface reconnect via interface currents carried along Fermi arcs. 
While the formation of chiral Landau levels is a purely quantum-mechanical effect, electron flow along the Fermi arcs can be understood semi-classically as a result of the Lorentz force changing the momentum by $|d\vk|/dt =  v/l_B^2$, where $l_B = \sqrt{\hbar/eB}\approx26$nm$\sqrt{1/B[\mathrm{T}]}$ is the magnetic length and $v$ is the Fermi-arc velocity. According to this equation of motion, the dwell time of a particle on a Fermi arc is 
\begin{equation}
    \td = \frac{\la l_B^2}{v} = \frac{\hbar\la}{evB}.
\end{equation}

For a clean interface, the conductance is simply that of the chiral (bulk) Landau level $G_0(B)$, which is perfectly transmitted via the Fermi arc, while contributions from higher Landau levels perfectly reflect \cite{Chaou2023}. In the presence of disorder, both effects are modified. The correction due to disorder-activated contribution of non-chiral Landau levels, however, stays negligible in most cases, as will be shown separately in Sec.\ \ref{sec:ncl}.
To get a sense of the effect of disorder on the transmission via the Fermi arcs, note that for well-separated Weyl nodes, $\la \sim 0.1\, \mathrm{\AA}^{-1}$, even at relatively large fields of $B\sim 1$T, the dwell time is on the order of $ \td \sim 1\mu\mathrm{m}/v$. Disorder introduces a finite lifetime $\tl$, which, depending on the purity of the interface, may well be shorter than $\td$. 
For this likely case of $\tl\lesssim\td$ one would expect particles to scatter within Fermi arcs (intra-arc scattering), as well as between the arcs (inter-arc scattering), the latter type of scattering event being between counter-propagating channels.

\subsection{Landauer-Boltzmann transport approach}

To calculate the conductance across the interface in the presence of disorder, we consider the in- and out-going chiral Landau levels in the bulk of the two WSMs, whose degeneracy is $N_B =(e/h)AB$, where $A B$ is the magnetic flux through the interface.
To linear order in the applied voltage, the non-equilibrium occupation of states $f$ can be expressed in terms of the deviation $\mu$ of the chemical potential from the Fermi level as $f=n_F+ n_F'\mu$, where $n_F$ is the equilibrium occupation and $n_F'$ its derivative with respect to energy. We denote these chemical-potential  deviations as $\mu_\pm(\kappa)$, where $\pm$ corresponds to the right/left-moving chiral states with chirality $\chi=\pm$  and $\kappa =\pm 1$ correspond to the right/left WSM. According to Landauer-B\"uttiker transport approach, the conductance is given by
\begin{equation}
     G(B) =  G_0(B)\, T(B), \label{eq:gB}
\end{equation}
where $G_0(B)$ is the conductance at full transmission of the chiral Landau levels and $T(B)$ is the transmission probability. Expressing the conductance in terms of the occupation functions $f=n_F+ n_F'\mu$, the transmission probability can be expressed in terms of $\mu_\pm(\kappa)$ as
\begin{equation}
  T(B) = \frac{\sum_\chi \chi \mu_\chi(\pm 1)}{\mu_+(-1)-\mu_-(1)}
\end{equation}
which is the occupation difference of countermoving modes (same in the right and left WSMs due to particle conservation) divided by the difference between left and right WSMs, $\mu_+(-1)-\mu_-(1)$, which corresponds to the applied voltage. 

To calculate the non-equilibrium potentials $\mu_{\pm}(\kappa)$ in the presence of disorder we consider the Boltzmann transport equation (BE), which describes the dynamics of a non-equilibrium occupation function $f$ of the Fermi arcs as
\begin{equation}
    \frac{\d f}{\d t} = \frac{d\vk}{dt}\cdot \partial_\vk f = \left( \frac{\partial f}{\partial t} \right)_\text{coll},
\end{equation}
where $d\vk/dt$ is given by the Lorentz force and the right hand side is the collision integral that perturbatively incorporates scattering processes. We used that $f$ is spatially homogeneous and $\partial f / \partial t = 0$ in the steady state. We use $f=n_F+ n_F'\mu$ and parametrize $\mu$ on the Fermi arc by allowing the parameter $\kappa$ in $\mu(\kappa)$ to assume values in the range $\kappa\in [-1,1]$, where $\kappa=\pm 1$ corresponds to the ends of the arc that merge with the bulk of the right/left WSM. 
Assuming zero temperature, approximating the collision integral to lowest order in the perturbation \cite{Kohn1957}, and using the Fermi-arc parametrization and Lorentz-force expression introduced above, the BE reads 
\begin{align}
        \chi  \partial_\kappa\mu_\chi (\kappa) = 
   \sum_{\chi '=\pm}\int_{-1}^1 d\kappa'\, \beta^{-1}_{\chi\kappa,\chi'\kappa'}[\mu_{\chi '} (\kappa')-\mu_\chi (\kappa)] 
\label{eq:BE}
\end{align}
where 
\begin{equation}
    \beta^{-1}_{\chi\kappa,\chi'\kappa'} =\beta^{-1}_{\chi'\kappa',\chi\kappa}= \frac{\la^2l_B^2}{v_\kappa v_{\kappa'}}\Gamma_{\chi'\kappa',\chi\kappa} 
\end{equation}
is the dwell-time-weighted scattering rate $\Gamma_{\chi'\kappa',\chi\kappa}$  (up to a constant) between the state at $(\chi\kappa)$ and $(\chi'\kappa')$, such that $\sum_{\chi'}\int d\kappa'\, \beta^{-1}_{\chi\kappa,\chi'\kappa'} =\td/\tl$.

\subsection{Approximate analytical solution}

Before we solve the BE in the low-field limit and (numerically) in an explicit microscopic model, we here consider an analytically tractable simplified model to better elucidate the magnetoconductance mechanism. We take the intra- and inter-arc scattering rates to assume two constant values, $\Gamma_{\chi\kappa,\chi'\kappa'}=\Gamma_\ia$ for $\chi=\chi'$ and $\Gamma_{\chi\kappa,\chi'\kappa'}=\Gamma_\ir$ for $\chi=-\chi'$. Solving for $\mu_\chi(\kappa)$ under the boundary condition of a given voltage across the interface, we obtain
\begin{equation}
    \mu_\chi(\kappa) = \frac{\chi}{2}\frac{e^{\frac{1-\chi\kappa}{2\beta}}(\gamma-1)+\gamma\beta\big(1-e^{\frac{1}{\beta}}\big)}{e^{\frac{1}{\beta}}(\gamma-1)+\gamma\beta\big(1-e^{\frac{1}{\beta}}\big)}
\end{equation}
leading to the transmission probability
\begin{align}
    T(B) =&\  1 - \frac12 \left( \frac{\beta\gamma}{1-\gamma} + \frac1{1 - e^{-1/\beta}}  \right)^{-1},
\end{align}
where
\begin{align}
    \gamma\equiv&\ \frac{\Gamma_\ia-\Gamma_\ir}{\Gamma_\ia+\Gamma_\ir},\ \\ 
    \beta\equiv&\ \frac{\td}{\tl}=\frac{\mfp}{\la l_B^2} = \frac{e|B|}{\hbar} \frac{\mfp}{\la}.
    \label{beta}
\end{align}

The parameter $\beta \sim 0.15 \, |B[\mathrm{T}]|\, \mfp[\mu\mathrm{m}]\, \la^{-1}[\mathrm{\AA}]$ is the flux through the area spanned by the mean free path $\mfp=\tl v$ and $\la^{-1}$ in units of the flux quantum $h/e$. For well-separated Weyl nodes, $\beta$ will be small in the experimentally relevant parameter regimes. The parameter $\gamma \in [-1,1]$ characterizes the dominant type of scattering. Spatially smooth (or correlated) disorder will tend to increase intra-arc scattering and lead to $\gamma\to 1$, while point-like (or uncorrelated) disorder would lead to similar scattering rates and thus $\gamma\to 0$. The regime $-1\leq\gamma\leq 0$ corresponds to dominant inter-arc scattering, which, though less realistic, is included for completeness.

\begin{figure}
	\includegraphics[width=\columnwidth]{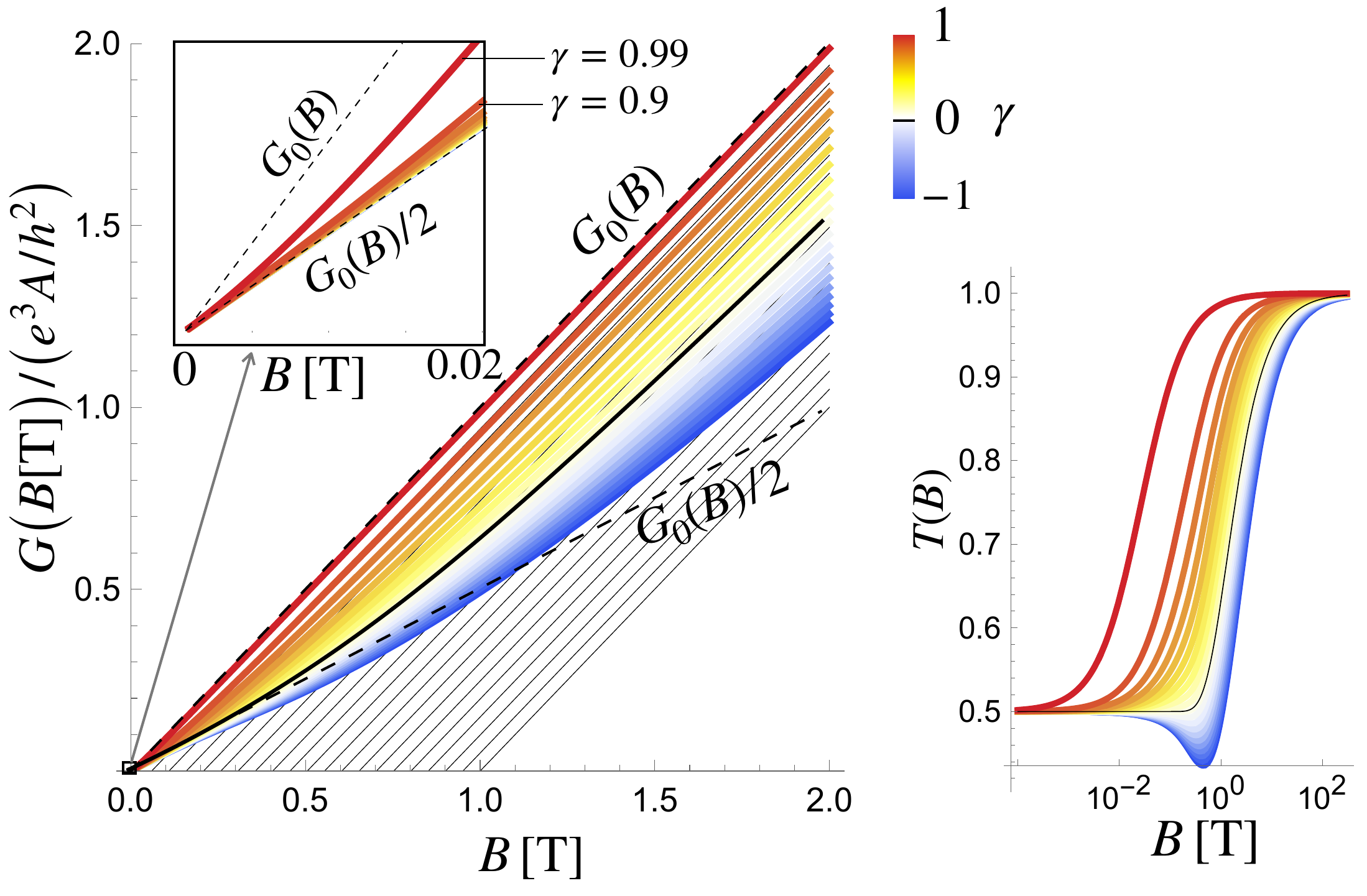}	
	\caption{Left: Conductance as a function of the magnetic field for $\gamma=0$ (black solid line) and all other $\gamma$ (colored lines). Dashed lines indicate the asymptotic conductance at low fields, $G_0(B)/2$, and large fields, $G_0(B)$ (thin gray lines for different $\gamma$). Upper inset shows a close-up at small-fields. Right: Transmission probability $T(B)=G(B)/G_0(B)$. The parameters are $\la=0.2 $\AA$^{-1}$, $\mfp=1\mu$m.	}
	\label{fig:Tplot}    
\end{figure}

The transmission probability is a non-analytic function of the field, whose behavior is shown in Fig.\ \ref{fig:Tplot}. For $0\leq \gamma \leq 1$ it increases monotonically from $0.5$ to $1$, while for (the less relevant regime) $-1\leq\gamma\leq 0$ it first slightly decreases to reach a minimum at $\beta\lesssim 0.3$ and $T(B) \gtrsim 0.4$. The limiting behavior near $T(0)$ and $T(\infty)$ is, however, $\gamma-$ independent. 

For small fields, the transmission saturates at $T(0)=\tfrac{1}{2}$; the asymptotic behavior at large fields is $T(B)|_{\beta\to\infty} = 1-\beta_0/\beta$, where $\beta_0 = (1-\gamma)/2$. This yields low- and high-field magnetoconductances of
\begin{align}
    G(B)=\begin{cases}
        G_0(B)/2 & \text{for }\ \  B\ll \ba \\
        G_0(B)-G_0(\ba) & \text{for }\ \  B\gg \ba,
    \end{cases}
\end{align}
where the characteristic field $\ba$ corresponding to $\beta_0$ reads
\begin{equation}
  \ba =\frac{\hbar}{e}\frac{1-\gamma}{2\mfp}\la  
  \equiv \frac{\hbar}{e}\frac{\la}{\lir} 
  \approx  \frac{3(1-\gamma)}{ \mfp [\mu\mathrm{m}]\, \la^{-1}[\mathrm{\AA}]}\mathrm{T},  
\end{equation}
where $\lir\equiv 2\mfp/(1-\gamma) = v/(\Gamma_\ir\la )$ is the mean free path between inter-arc scattering events and thus the characteristic length for which a particle does not change the arc. Note that  $\tau_\mathrm{arc}\equiv\lir/v$ can be interpreted as a Fermi-arc lifetime and  $\ba$ as the characteristic field at which the arc dwell time is equal to the arc lifetime, $\td=\tau_\mathrm{arc}$; it is  independent of $\Gamma_\ia$. 

\subsection{Universal low-field limit}

An analytically tractable limit of the full BE is the low-field regime, $\td\gg\tl$, which we discuss in the following without  limiting ourselves to the simplified model. We consider $N_L$ positive-chirality modes in the left WSM that move to the right, all at the occupation
	$\mu_+$, and $N_R$ negative-chirality modes in the right WSM that move to the left, all at the occupation
	$\mu_-$. The conductance  generalized to this  case is given by
	\begin{equation}
		G(B) =   \frac{\sum_a \chi_a \mu_a(\pm 1)}{\mu_+-\mu_-}G_0(B), \label{cn2}
	\end{equation}
where the sum runs over all chiral modes, and the BE generalizes to 
	\begin{align}
	  	\chi_a \partial_\kappa\mu_a (\kappa) =
		\sum_{a'}\int d\kappa'\, \beta^{-1}_{a\kappa,a'\kappa'}[\mu_{a '} (\kappa')-\mu_a(\kappa)]. 
		\label{eq:BE2}
	\end{align}
	
	In the limit of dwell time being much larger than the lifetime,  $\beta_{a\kappa,a'\kappa'} \to 0 $, the solution assumes the form
\begin{align}
	\mu_a(\kappa) = e^{-(1+\chi_a\kappa)/\kappa_a} \mu_{\chi_a}+\epsilon_{a}(\kappa)
\end{align}
where  $\epsilon_{a}(\kappa)\sim\kappa_a
\sim \beta\to 0$, which can be proven straight forwardly by inserting into the BE. Hence, in this limit, the occupations fall to the equilibrium values (zero) quickly when propagating along the arc, so that $\mu_a(1)=0$ for $\chi_a=+$ and $\mu_a(-1)=0$ for $\chi_a=-$.  Equating the conductance calculated in the left and right lead in Eq.\ \eqref{cn2}, one obtains 
$N_\L \mu_+ +N_\R\mu_-=0$. Inserting back into \eqref{cn2} gives the result
\begin{equation}
		G(B)  = \frac{N_\L N_\R}{N_\L+N_\R} G_0(B).
\end{equation}
In the low-field limit, the conductance thus assumes an universal fraction of $G_0(B)$, given by the numbers of Weyl nodes in the two WSMs.

\subsection{Summary and interpretation of the analytical results}

We first note that since only inter-arc scattering leads to current relaxation, transmission through the interface is unaffected by disorder if the Fermi-arc lifetime --- the time between \textit{inter-arc} scattering events --- is much longer than the dwell time, $\tau_\mathrm{arc} \gg \td \propto 1/B$. 

In the limit of  large magnetic fields we thus obtain
for a minimal model of a single pair of homochiral Fermi arcs $G(B)=G_0(B)-G_0(\ba)$, where the slope $dG(B)/dB$ is identical to the universal clean-interface case. Interpolation of the linear magnetoconductance at high fields towards zero field, can give a characteristic offset $G(0)=-G_0(\ba)$ that depends on the Fermi-arc- length and lifetime. 
For an arbitrary interface the clean-interface conductance is given by $N_\mathrm{ho} G_0(B)$, where $N_\mathrm{ho}$ is the number of homochiral Fermi-arc pairs.

\begin{figure*}
	\includegraphics[width=\textwidth]{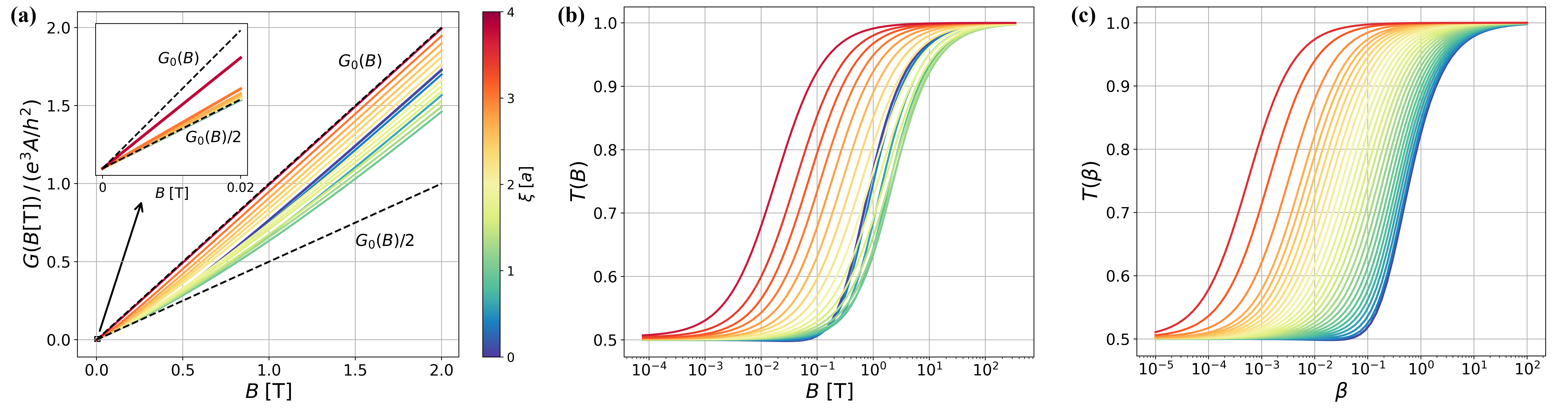}	
	\caption{(a) Conductance $G(B)$ for the WSM model Eq. \eqref{eq:hlt_orig} with random correlated disorder Eq. \eqref{eq:dis} based on the hybrid Landauer-Boltzmann approach for different disorder correlation lengths $\xi$ (cf.\ Fig.\ \ref{fig:Tplot}). (b) Transmission $T(B)=G(B)/G_0(B)$ as a function of $B$ and (c) as a function of $\beta$. Model parameters are $t=0.4$ ($\Delta\approx 1/a$), $\la=0.2/a $, $\mfp=10^4a$, $a=1$\AA. }
	\label{fig:BEplot}    
\end{figure*}

In the weak-field limit, i.e., $\tau_\mathrm{arc} \ll \td$, scattering between Fermi arcs leads to an equalized probability of exiting the interface via any of the $N_\L$+$N_\R$ chiral bulk modes in the left and right WSM, leading to a universal low-field magnetoconductance $G(B)/G_0(B) =N_\L N_\R/(N_\L+N_\R)$ (which is $1/2$  in the minimal model of two Weyl nodes). This is a momentum-space analog of the fractionally quantized conductance across a graphene p-n junction in the quantum Hall regime, where the equilibration of co-propagating edge states along the junction in \textit{real} space similarly leads to scattering into all outgoing modes with equal probability and hence, a fractionally quantized conductance \cite{Williams2007,Abanin2007}. 

In the intermediate regime of $\tau_\mathrm{arc} \sim \td$, the transmission probability changes in a model-detail-dependent way. In the minimal model of a singel pair of Weyl nodes, when inter-arc (intra-arc) scattering dominates, it becomes more likely for a particle to exit on the opposite (same) Fermi arc, thereby giving a conductance slightly smaller (larger) than $G_0(B)/2$. 

\section{Numerical simulations}
\label{sec:numerics}

We now substantiate the discussion of the previous section by numerical computations on an explicit lattice model.

\subsection{Lattice model }
\label{sec:kwant}

We consider two WSMs labeled by $A \in \{\L, \R\}$, described by the Bloch Hamiltonians
\begin{align}
	\hlt_A(k_x,\vk) &=  \sin k_x \tau^x  + \eta_y^A(\ktrans) \tau^y + (1 + 2\cos k_0 \nonumber \\ 
	&\; - \cos k_x - \cos k_y - \cos k_z) \tau^z, 
	\label{eq:hlt_orig} 
\end{align}
where Pauli matrices $\tau^a$ represent a pseudospin degree of freedom, $\vk=(k_y,k_z)$, and the lattice constant is set to unity (but is reinserted where necessary). The parameter $k_0$ determines the positions of the Weyl nodes, which lie in the $k_x = 0$ plane at $\ktrans_\L = \pm(k_0, k_0)$ for $A=\L$ and $\ktrans_\R = \pm(k_0, -k_0)$ for $A=\R$. Accordingly,
\begin{align} 
	\eta_y^\L(\vk) &= \sin{k_0} ( \sin k_y - \sin k_z) /{\sin{|\ktrans_\L|}} , \\[2pt]
	\eta_y^\R(\vk) &= \sin{k_0} ( \sin k_y + \sin k_z)/{\sin{|\ktrans_\R|}}.
\end{align}

An interface at $x=0$ between the WSMs $A=\L$ to the left and $A=\R$ to the right is introduced by transformation into real space in $x$ direction and reducing the hopping  across the interface by the factor $\tu\in [0,1]$. 
For decoupled WSMs ($\tu=0$) Fermi arcs given by $\eta_y^A(\ktrans) = 0$ are straight perpendicular lines joining the nodes of the corresponding WSM and crossing each other near $\vk=\nullv$ (at low energy). 
Turning on the interface coupling $\tu$, the Fermi arcs hybridize and the crossing becomes an anticrossing with a minimum separation $\Delta$ between the arcs, as illustrated in Fig.\ \ref{fig:1}.

We model disorder by a random onsite potential on the interface, $V(\vr)$, with 
\begin{equation}
    \braket{V(\vr)} = 0, \quad 
    \braket{V(\vr)V(\vr')} = W^2 f(\vr-\vr'),  \label{eq:dis}
\end{equation}
where $\vr=(y,z)$ is the position on the interface, $\braket{\dots}$ denotes average over disorder configurations, $W$ parametrizes the magnitude of the disorder potential and $f(\vr-\vr')$ encodes the disorder autocorrelation. We choose  
\begin{equation}
  f(\vr) = \frac{1}{\pi\xi^2} \e^{-\abs{\vr}^2/\xi^2}, 
\end{equation}
where $\xi$ is the disorder correlation length.  Uncorrelated disorder is obtained in the limit $\xi\to 0$. 
	
We numerically compute the conductance across such an interface using \textit{Kwant} python package \cite{Groth2014}. We consider a system with a cross section of $40\times40$ unit cells in the $y$ and $z$ directions, with periodic boundary conditions. The scattering region consists of the two interface layers of the right/left WSM as given by the Eq. \eqref{eq:hlt_orig} but with the interface hopping suppressed by the factor $\tu$. Each interface layer is attached to right/left lead, consisting of the right/left WSM from Eq.\ \eqref{eq:hlt_orig}.  Onsite Anderson disorder is added to each site in the scattering region, sampled from a distribution that is taken to be either a Gaussian of width $W$ and correlation length $\xi$ or, for uncorrelated disorder, a uniform distribution on $(-W\sqrt3, W\sqrt3)$ (such that the variance is $W^2$). The conductance is averaged over 300 disorder realizations. The results of the numerical simulations will be discussed below, in comparison with those of the Landauer-Boltzmann approach.

\subsection{Landauer-Boltzmann approach}
\label{sec:LB_approach}

For the correlated-disorder model, the scattering rate in the BE \eqref{eq:BE} is given by \cite{Kohn1957}
\begin{equation}
	\Gamma_{\vk,\vk'} = W^2 f(\vk-\vk')\abs{\braket{\Psi_{\vk} | \Psi_{\vk'}}}^2, \label{eq:scatrate}
\end{equation}
where $f(\vk)=e^{-\vk^2\xi^2/4}/\xi\pi\sqrt{2}$ is the Fourier-transformed correlation function $f(\vr)$ and $\ket{\Psi_{\vk}}$ are Fermi-arc states. 

For correlated disorder ($\xi > 0$), the typical momentum change upon scattering is limited by the inverse correlation length $\abs{\vk-\vk'}\lesssim 1/\xi$. Inter-arc scattering is therefore strongly suppressed if $1/\xi$ is smaller than the Fermi arc separation $\Delta$. Otherwise, for $1/\xi\gtrsim \Delta$, both inter- and intra-arc scattering should be present. Thus, comparing with the analytic solution in Fig.\ \ref{fig:Tplot}, we expect $\xi\gtrsim 1/\Delta$ to correspond to the limit $\gamma\to 1$ and $\xi\lesssim 1/\Delta$ to  the limit $\gamma\to 0$. As mentioned above, we do not expect to find dominant inter-arc scattering (\textit{i.e.}, $-1<\gamma<0$) for any $\xi$ and thus we expect the transmission to lie between $0.5$ and $1$.

To calculate the conductance, we solve the BE \eqref{eq:BE} for the microscopic model Eq.\ \eqref{eq:hlt_orig}. The dwell time $\td^{-1}=v/\la l_B^2$ in the BE is momentum dependent due to the (weak) momentum dependence of the velocity $v$. The calculation of the velocity and the spinor overlap $\braket{\Psi_{\vk} | \Psi_{\vk'}}$ is detailed in Appendix~\ref{app:tmat}. The BE is solved numerically via a discretization of the arcs in momentum space. 

The resulting conductance  $G(B)$ and transmission $T(B)=G(B)/G_0(B)$ are plotted in Fig.\ \ref{fig:BEplot} for different correlation lengths $\xi$. The parameters in Fig.\ \ref{fig:BEplot} give a characteristic crossover field of $\ba\approx (1-\gamma)$T and a Fermi-arc separation $\Delta\approx1/a=1$\AA$^{-1}$. The plots confirm the behavior anticipated from the analytic solution: The transmission lies within the expected range of $0.5 < T(B) < 1$, whereby $T(B)=1$ for $B\gg \ba$ and $T(B)=1/2$ for $B\gg \ba$. Regarding the dependence on $\xi$, we find that $\ba$ goes to zero for $\xi\gg 1/\Delta$, as expected from the exponentially suppressed inter-arc scattering ($\gamma\to1$). 

\begin{figure}
    \centering

    \raggedright{\scriptsize \textbf{\hspace*{0.5em}(a)}}\\[0.2em] 
    \includegraphics[width=\columnwidth]{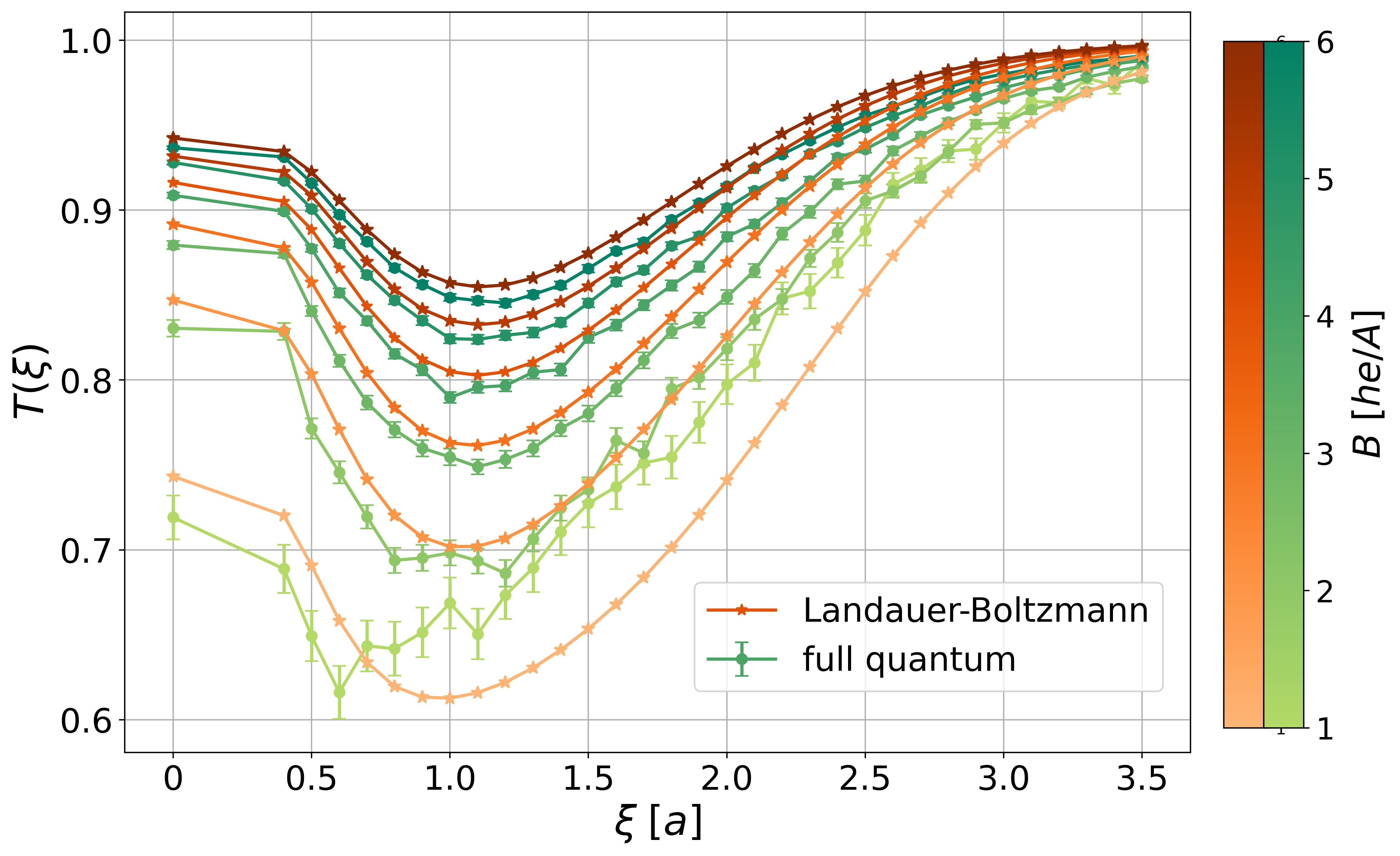}
    \raggedright{\scriptsize \textbf{\hspace*{0.5em}(b)}}\\[0.2em]
    \includegraphics[width=\columnwidth]{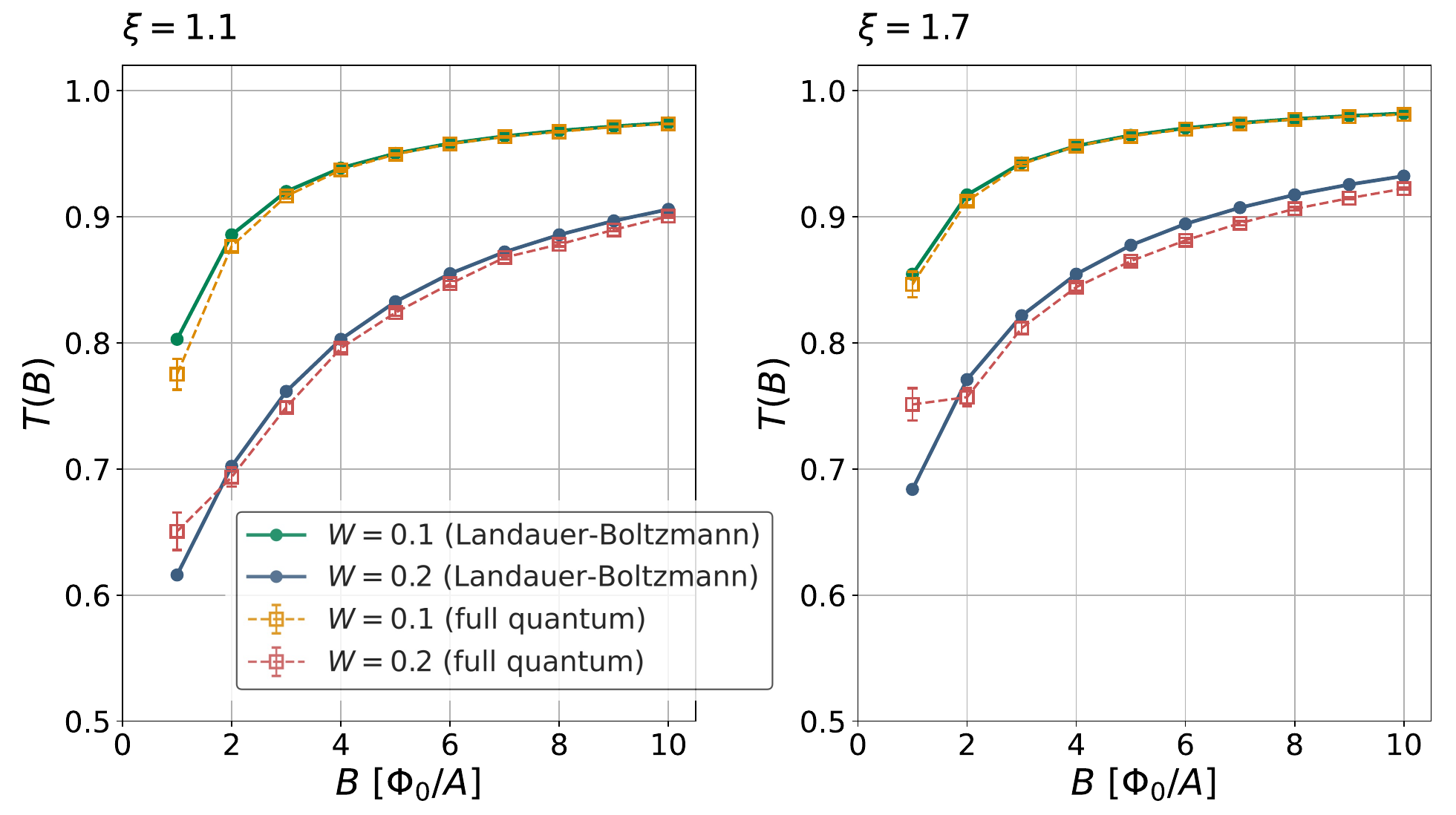}
    
	\caption{(a) Transmission $T(B)=G(B)/G_0(B)$ as a function of the scattering correlation length $\xi$ for different magnetic field strength ($B[h/eA]$ is the number of flux quanta through the interface area $A$). The plot shows results from the Landauer-Boltzmann approach and the full-quantum simulation of a $40\times40$ interface. Model parameters are $t=0.4$, $W=0.2$, and energy $E=0.1$. (b) Transmission as a function of $B$ for two different $\xi$ and $W$ values. Other parameters as in (a).}
	\label{fig:kwant}    
\end{figure}

Note that due to the additional dependence of the scattering rate \eqref{eq:scatrate} on the wavefunction overlap, the ratio of life and dwell times (parameter $\beta$ in the simplified model)  depends weakly on $\xi$. As a result, in the crossover regime $\xi\sim 1/\Delta$, the $\xi$ dependence of the transmission at a fixed field shows a non-monotic  behavior, visible more clearly in Fig.\ \ref{fig:kwant} discussed below. To make a better comparison with the analytic solution, we determine the life and dwell times for the lattice model and plot the transmission as a function of $\beta$, shown in Figure \ref{fig:BEplot}(c). The plot clearly shows that, indeed, for a fixed $\beta$, the transmission increases monotonically with $\xi$, as expected (c.f.\ Fig.\ \ref{fig:Tplot} right).

Finally, in Fig.\ \ref{fig:kwant}, we plot the transmission $T(B) = G(B)/G_0(B)$ (where $G_0(B)$ is the theoretical clean-interface conductance \eqref{cond0}) as a function of the scattering correlation length $\xi$ for the full quantum simulation as well as the Landauer-Boltzmann approach. The two results are in agreement (without any fitting parameter) even for a relatively large disorder strength $W=0.2$, and the agreement improves further with decreasing $W$ [Fig.\ \ref{fig:kwant}(b)]. The largest deviation at small fields is attributed to small corrections arising from disorder-activated contributions of non-chiral Landau levels, discussed further below. 

\subsection{Magnetic breakdown at weakly coupled interfaces} \label{sec:mb}

In the above analysis, we assumed semiclassical transport along the interface Fermi arcs, which is justified if the arcs are always separated by a distance larger than $l_B^{-1}\approx 0.004\mathrm{\AA}^{-1}\sqrt{B[\mathrm{T}]}$. 
Since this distance is very small at realistic magnetic fields, the above results apply to most realistic interfaces. 

At an interface with a particularly weak tunnel coupling, which may be purposefully designed by encapsulating an insulating layer between the WSMs, the semiclassics breaks down. Here, the minimal separation at the avoided crossing of the decoupled Fermi arcs is set by the weak coupling across the interface and can become sufficiently small. In this case, particles can quantum tunnel between the arcs --- a phenomenon called magnetic breakdown \cite{Chaou2023}. As a result, the transmission probability is suppressed, which in the clean case has been quantified as \cite{Chaou2023} 
\begin{equation}
	T_\mathrm{MB}(B) =  1-\e^{-B_0/B}, \quad B_0 = \frac\pi4 \Delta^2 \tan\theta,  
\end{equation}
where $B_0$ is the breakdown field that depends only on the minimum separation $\Delta$ (see Fig.\ \ref{fig:1}) and the opening angle $2\theta$ of the arcs.
The suppressed transmission leads to a saturation of the conductance, $G(B\to\infty)\to G(B_0)$.

\begin{figure}	
\includegraphics[width=\columnwidth]{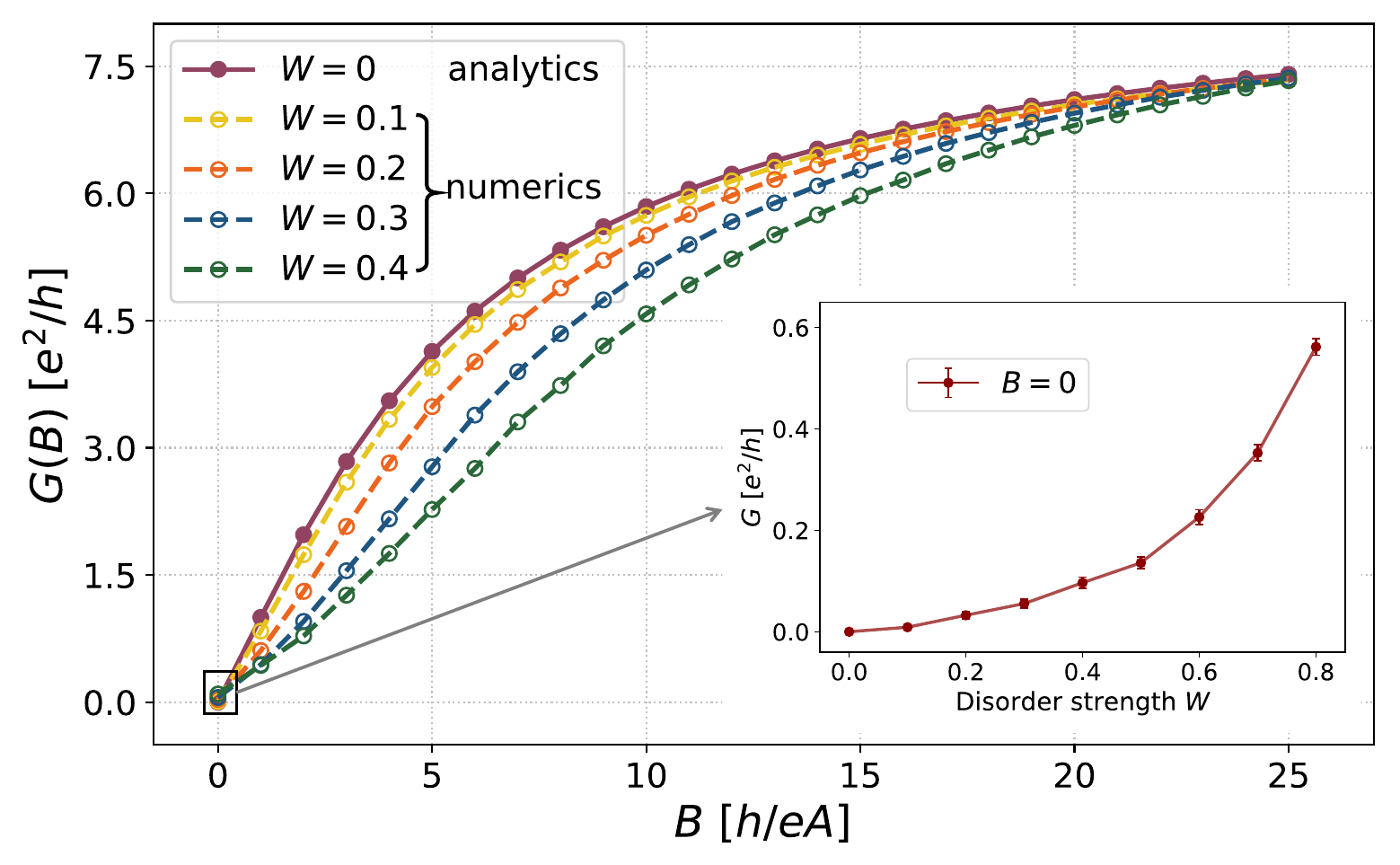}	
	\caption{Conductance of weakly coupled interfaces (exhibiting magnetic breakdown) for different (uncorrelated) disorder strength $W$ at energy $E=0.1$ and interface hopping strength $t=0.1$. Disorder shows a weak suppression of the conductance, similar to the normal-interface case. Weak disorder-activated conductance of non-chiral Landau levels is visible as a small zero-field conductance, shown in the inset.
    }
	\label{fig:mbkwant}    
\end{figure}

The analytical Landauer-Boltzmann approach does not capture the effect of magnetic breakdown since the motion along the arcs is treated semiclassically. The numerical simulation for a weakly coupled, disordered interface is shown in Fig.\ \ref{fig:mbkwant}. 
The comparison with the clean case shows that the disorder only weakly modifies the qualitative behavior dictated by magnetic breakdown, with disorder effects being negligible in the large field regime ($B \gg B_0$). 

\subsection{Contributions from non-chiral Landau levels}
\label{sec:ncl}

The universality of the magnetoconductance through a clean interface is partly due to the fact that all non-chiral Landau levels perfectly reflect at the interface \cite{Chaou2023}. This is no longer strictly true if the momentum parallel to the interface is not conserved, as happens in the presence of disorder. We will now show that these corrections are negligible. 

While we neglect the contributions of non-chiral Landau levels in the Landauer-Boltzmann approach, such contributions can in principle show up in the numerical transport simulations.  In Fig.\ \ref{fig:kwant} we indeed observe a weak enhancement at small fields and intermediate $\xi$. In Fig.\ \ref{fig:mbkwant} the enhancement is also visible as a tiny zero-field conductance that increases with an increasing disorder strength. 

The weakness of these corrections can be explained as follows. The number of (non-chiral) bulk modes that arrive at the interface area $A$ is at most  on the order of $\sim k_F^2A$, where $k_F\ll \la$ is the Fermi momentum measured from the Weyl node. This number decreases with increasing $B$ due to an increased occupation of the zeroth Landau level. Since the overlap of bulk states of right and left WSM is much smaller than that of bulk and Fermi arcs, the rate of transmission of non-chiral bulk states is on the order of the scattering rate between the bulk and the Fermi arcs. The disorder-activated conductance can thus be estimated as $(e^2/h) k_F A^2/\mfp \la$. Comparing with the anomalous magnetoconductance $G(B)\sim G_0(B)$ in Eq. \eqref{cond0} we find that the disorder-induced contribution of non-chiral levels becomes significant only if 
\begin{equation}
    \frac{1}{l_B^2} \lesssim \frac{k_F^2}{\mfp \la}.
\end{equation}
This explains why there is no significant effect in the case of weak disorder and well-separated Weyl nodes.

\section{Discussion and conclusions}
\label{sec:conc}

We have explored the robustness of the tunnel magnetoconductance across a WSM interface, which in the ideal limit of a clean interface is given by $G_0(B)=(e^3/h^2) \abs{\boldsymbol{A}\cdot \boldsymbol{B}} $ for each homochiral (i.e., chirality-preserving) Fermi arc, where $\abs{ \boldsymbol{A}\cdot \boldsymbol{B}}$ is the magnetic flux through the interface of area $\boldsymbol{A}$. Our results show that disorder introduces two  field regimes, above and below a characteristic field strength $B\approx\ba$, at which the Fermi-arc lifetime $\tau_\mathrm{arc}$---the mean time between inter-arc scattering events---is equal to the dwell time $\td$---the time a particle spends on the Fermi arc in the presence of a magnetic field. When the inter-arc scattering is dominated by spatially smooth disorder, $\ba$ goes to zero as $\sim e^{-\Delta^2\xi^2}$ with an increasing disorder correlation length $\xi$ and an increasing separation of the Fermi arcs $\sim\Delta$.

In the low-field regime, $\tau_\mathrm{arc}\ll\td$,
for a system with $N_\L$ ($N_\R$) Weyl node pairs in the left (right) WSM, the low-field magnetoconductance will assume a simple fraction of the elementary  conductance,
\begin{equation}
    G(B) =\frac{N_\L N_\R}{N_\L+N_\R} G_0(B),
\end{equation}
independent on the Fermi-arc connectivity. This result resembles the fractionally quantized conductance across a graphene p-n junction in the quantum Hall regime, where the role of motion along Fermi arcs in \textit{momentum} space is played by motion along co-propagating edge states along the junction in \textit{real} space \cite{Williams2007,Abanin2007}.

In the high-field regime ($\tau_\mathrm{arc}\gg\td$), we recover the clean-interface magnetoconductance. Here the crucial numbers are not the numbers of Weyl nodes but instead the number of \textit{homochiral} Fermi arc pairs $N_\mathrm{ho}$ --- Fermi arcs that connect projections of Weyl nodes of the same chirality, which are then necessarily Weyl nodes from opposite sides of the interface. The magnetoconductance is then equal to
\begin{equation}
    G(B)=N_\mathrm{ho}G_0(B).
\end{equation}
The three integer parameters $N_\mathrm{ho}$, $N_\L$, and $N_\R$ can be independently controlled by the choice of the two interfaced WSM materials and the interface design. In this way, various differently shaped magnetoconductance behavior can be experimentally realized.

To test and exemplify these predictions, we considered
a minimal model of a single pair of Weyl nodes, which position shifts across the interface and the four node projections are connected by two homochiral Fermi arcs. The resulting magnetoconductance behavior, in particular showing a low- and high-field conductance of $G(B)=G_0(B)/2$ and $G(B)=G_0(B)$, respectively, was confirmed by comparing with numerical simulations on a lattice model. 

In the experimental measurement of the longitudinal magnetoresistance in a grained WSM \cite{Zhang2023} with randomly oriented crystallites, a robust negative linear magnetoresistance has been observed across a large field range. The slope at fields below $\sim 1$T is approximately a factor of $2$ smaller than the slope at larger fields. At fields far below the saturation field (no saturation is observed in the experiment), the corresponding magnetoconductance is minus the magnetoresistance, which thus qualitatively agrees with our theory. Since the number of Weyl nodes does not change across the graines, the slope difference by a factor two indicates, according to our analysis, that the interface Fermi arcs are predominantly homochiral --- something that is interesting to check by future material-science work.

Methodologically, we derived a semiclassical Landauer-Boltzmann theory, whose validity we confirmed by comparing with numerical transport simulations of a disordered system with both correlated and uncorrelated disorder at zero temperature. We expect the semiclassical theory to remain valid at finite temperatures, as well as for other perturbations, such as elastic and in-elastic scattering off phonons. Our hybrid Landauer-Boltzmann theory can be easily extended to various scenarios to gain analytical insight or more efficient numerics (as compared to full-quantum calculations).

Qualitatively, we can infer that the temperature dependence of the anomalous magnetoconductance contribution is weak, since the energy dependence (in the range of separated Weyl fermions) is vanishing/weak. The energy dependence vanishes in the universal low- and high-field limits as the clean-system conductance $G_0(B)$ is energy independent \cite{Chaou2023}. In the crossover regime, the conductance could in general depend on energy through the energy dependence of the scattering and the dwell time but the conductance can only be modified by a strongly bounded factor $\in[0.5,1]$. 

\begin{acknowledgments}
We thank Piet Brouwer and Zhiyang Tan for useful discussions. This research was funded by the Deutsche Forschungsgemeinschaft (DFG, German Research Foundation) through CRC-TR 183 “Entangled States of Matter” and the Emmy Noether program, Project No. 506208038.
\end{acknowledgments}

\bibliography{library}	

\providecommand{\noopsort}[1]{}
\begin{thebibliography}{32}%
\makeatletter
\providecommand \@ifxundefined [1]{%
 \@ifx{#1\undefined}
}%
\providecommand \@ifnum [1]{%
 \ifnum #1\expandafter \@firstoftwo
 \else \expandafter \@secondoftwo
 \fi
}%
\providecommand \@ifx [1]{%
 \ifx #1\expandafter \@firstoftwo
 \else \expandafter \@secondoftwo
 \fi
}%
\providecommand \natexlab [1]{#1}%
\providecommand \enquote  [1]{``#1''}%
\providecommand \bibnamefont  [1]{#1}%
\providecommand \bibfnamefont [1]{#1}%
\providecommand \citenamefont [1]{#1}%
\providecommand \href@noop [0]{\@secondoftwo}%
\providecommand \href [0]{\begingroup \@sanitize@url \@href}%
\providecommand \@href[1]{\@@startlink{#1}\@@href}%
\providecommand \@@href[1]{\endgroup#1\@@endlink}%
\providecommand \@sanitize@url [0]{\catcode `\\12\catcode `\$12\catcode
  `\&12\catcode `\#12\catcode `\^12\catcode `\_12\catcode `\%12\relax}%
\providecommand \@@startlink[1]{}%
\providecommand \@@endlink[0]{}%
\providecommand \url  [0]{\begingroup\@sanitize@url \@url }%
\providecommand \@url [1]{\endgroup\@href {#1}{\urlprefix }}%
\providecommand \urlprefix  [0]{URL }%
\providecommand \Eprint [0]{\href }%
\providecommand \doibase [0]{https://doi.org/}%
\providecommand \selectlanguage [0]{\@gobble}%
\providecommand \bibinfo  [0]{\@secondoftwo}%
\providecommand \bibfield  [0]{\@secondoftwo}%
\providecommand \translation [1]{[#1]}%
\providecommand \BibitemOpen [0]{}%
\providecommand \bibitemStop [0]{}%
\providecommand \bibitemNoStop [0]{.\EOS\space}%
\providecommand \EOS [0]{\spacefactor3000\relax}%
\providecommand \BibitemShut  [1]{\csname bibitem#1\endcsname}%
\let\auto@bib@innerbib\@empty
\bibitem [{\citenamefont {Wan}\ \emph {et~al.}(2011)\citenamefont {Wan},
  \citenamefont {Turner}, \citenamefont {Vishwanath},\ and\ \citenamefont
  {Savrasov}}]{Wan2011}%
  \BibitemOpen
  \bibfield  {author} {\bibinfo {author} {\bibfnamefont {X.}~\bibnamefont
  {Wan}}, \bibinfo {author} {\bibfnamefont {A.~M.}\ \bibnamefont {Turner}},
  \bibinfo {author} {\bibfnamefont {A.}~\bibnamefont {Vishwanath}},\ and\
  \bibinfo {author} {\bibfnamefont {S.~Y.}\ \bibnamefont {Savrasov}},\
  }\bibfield  {title} {\bibinfo {title} {Topological semimetal and
  {{Fermi-arc}} surface states in the electronic structure of pyrochlore
  iridates},\ }\href {https://doi.org/10.1103/PhysRevB.83.205101} {\bibfield
  {journal} {\bibinfo  {journal} {Physical Review B}\ }\textbf {\bibinfo
  {volume} {83}},\ \bibinfo {pages} {205101} (\bibinfo {year}
  {2011})}\BibitemShut {NoStop}%
\bibitem [{\citenamefont {Armitage}\ \emph {et~al.}(2018)\citenamefont
  {Armitage}, \citenamefont {Mele},\ and\ \citenamefont
  {Vishwanath}}]{Armitage2018}%
  \BibitemOpen
  \bibfield  {author} {\bibinfo {author} {\bibfnamefont {N.~P.}\ \bibnamefont
  {Armitage}}, \bibinfo {author} {\bibfnamefont {E.~J.}\ \bibnamefont {Mele}},\
  and\ \bibinfo {author} {\bibfnamefont {A.}~\bibnamefont {Vishwanath}},\
  }\bibfield  {title} {\bibinfo {title} {Weyl and {{Dirac}} semimetals in
  three-dimensional solids},\ }\href
  {https://doi.org/10.1103/RevModPhys.90.015001} {\bibfield  {journal}
  {\bibinfo  {journal} {Reviews of Modern Physics}\ }\textbf {\bibinfo {volume}
  {90}},\ \bibinfo {pages} {015001} (\bibinfo {year} {2018})}\BibitemShut
  {NoStop}%
\bibitem [{\citenamefont {Adler}(1969)}]{Adler1969}%
  \BibitemOpen
  \bibfield  {author} {\bibinfo {author} {\bibfnamefont {S.~L.}\ \bibnamefont
  {Adler}},\ }\bibfield  {title} {\bibinfo {title} {Axial-vector vertex in
  spinor electrodynamics},\ }\href@noop {} {\bibfield  {journal} {\bibinfo
  {journal} {Physical Review}\ }\textbf {\bibinfo {volume} {177(5)}},\ \bibinfo
  {pages} {2426} (\bibinfo {year} {1969})}\BibitemShut {NoStop}%
\bibitem [{\citenamefont {Bell}\ and\ \citenamefont {Jackiw}(1969)}]{Bell1969}%
  \BibitemOpen
  \bibfield  {author} {\bibinfo {author} {\bibfnamefont {J.~S.}\ \bibnamefont
  {Bell}}\ and\ \bibinfo {author} {\bibfnamefont {R.}~\bibnamefont {Jackiw}},\
  }\bibfield  {title} {\bibinfo {title} {A {{PCAC Puzzle}}: {$\pi^0 \rightarrow
  \gamma\gamma$} in the {$\sigma$}-model},\ }\href
  {https://doi.org/10.1007/BF02823296} {\bibfield  {journal} {\bibinfo
  {journal} {Il Nuovo Cimento A (1965-1970)}\ }\textbf {\bibinfo {volume}
  {60}},\ \bibinfo {pages} {47} (\bibinfo {year} {1969})}\BibitemShut {NoStop}%
\bibitem [{\citenamefont {Son}\ and\ \citenamefont {Spivak}(2013)}]{Son2013a}%
  \BibitemOpen
  \bibfield  {author} {\bibinfo {author} {\bibfnamefont {D.~T.}\ \bibnamefont
  {Son}}\ and\ \bibinfo {author} {\bibfnamefont {B.~Z.}\ \bibnamefont
  {Spivak}},\ }\bibfield  {title} {\bibinfo {title} {Chiral anomaly and
  classical negative magnetoresistance of {{Weyl}} metals},\ }\href
  {https://doi.org/10.1103/PhysRevB.88.104412} {\bibfield  {journal} {\bibinfo
  {journal} {Physical Review B}\ }\textbf {\bibinfo {volume} {88}},\ \bibinfo
  {pages} {104412} (\bibinfo {year} {2013})}\BibitemShut {NoStop}%
\bibitem [{\citenamefont {Burkov}(2018)}]{Burkov2017}%
  \BibitemOpen
  \bibfield  {author} {\bibinfo {author} {\bibfnamefont {A.~A.}\ \bibnamefont
  {Burkov}},\ }\bibfield  {title} {\bibinfo {title} {{Weyl Metals}},\ }\href
  {https://doi.org/10.1146/annurev-conmatphys-033117-054129} {\bibfield
  {journal} {\bibinfo  {journal} {Annu. Rev. Condens. Matter Phys.}\ }\textbf
  {\bibinfo {volume} {9}},\ \bibinfo {pages} {359} (\bibinfo {year}
  {2018})}\BibitemShut {NoStop}%
\bibitem [{\citenamefont {Xiong}\ \emph {et~al.}(2015)\citenamefont {Xiong},
  \citenamefont {Kushwaha}, \citenamefont {Liang}, \citenamefont {Krizan},
  \citenamefont {Hirschberger}, \citenamefont {Wang}, \citenamefont {Cava},\
  and\ \citenamefont {Ong}}]{Xiong2015}%
  \BibitemOpen
  \bibfield  {author} {\bibinfo {author} {\bibfnamefont {J.}~\bibnamefont
  {Xiong}}, \bibinfo {author} {\bibfnamefont {S.~K.}\ \bibnamefont {Kushwaha}},
  \bibinfo {author} {\bibfnamefont {T.}~\bibnamefont {Liang}}, \bibinfo
  {author} {\bibfnamefont {J.~W.}\ \bibnamefont {Krizan}}, \bibinfo {author}
  {\bibfnamefont {M.}~\bibnamefont {Hirschberger}}, \bibinfo {author}
  {\bibfnamefont {W.}~\bibnamefont {Wang}}, \bibinfo {author} {\bibfnamefont
  {R.~J.}\ \bibnamefont {Cava}},\ and\ \bibinfo {author} {\bibfnamefont
  {N.~P.}\ \bibnamefont {Ong}},\ }\bibfield  {title} {\bibinfo {title}
  {{Evidence for the chiral anomaly in the Dirac semimetal Na$_3$Bi}},\ }\href
  {https://doi.org/10.1126/science.aac6089} {\bibfield  {journal} {\bibinfo
  {journal} {Science}\ }\textbf {\bibinfo {volume} {350}},\ \bibinfo {pages}
  {413} (\bibinfo {year} {2015})}\BibitemShut {NoStop}%
\bibitem [{\citenamefont {Altland}\ and\ \citenamefont
  {Bagrets}(2016)}]{Altland2016}%
  \BibitemOpen
  \bibfield  {author} {\bibinfo {author} {\bibfnamefont {A.}~\bibnamefont
  {Altland}}\ and\ \bibinfo {author} {\bibfnamefont {D.}~\bibnamefont
  {Bagrets}},\ }\bibfield  {title} {\bibinfo {title} {Theory of the strongly
  disordered {{Weyl}} semimetal},\ }\href
  {https://doi.org/10.1103/PhysRevB.93.075113} {\bibfield  {journal} {\bibinfo
  {journal} {Physical Review B}\ }\textbf {\bibinfo {volume} {93}},\ \bibinfo
  {pages} {75113} (\bibinfo {year} {2016})}\BibitemShut {NoStop}%
\bibitem [{\citenamefont {dos Reis}\ \emph {et~al.}(2016)\citenamefont {dos
  Reis}, \citenamefont {Ajeesh}, \citenamefont {Kumar}, \citenamefont {Arnold},
  \citenamefont {Shekhar}, \citenamefont {Naumann}, \citenamefont {Schmidt},
  \citenamefont {Nicklas},\ and\ \citenamefont {Hassinger}}]{Reis2016}%
  \BibitemOpen
  \bibfield  {author} {\bibinfo {author} {\bibfnamefont {R.~D.}\ \bibnamefont
  {dos Reis}}, \bibinfo {author} {\bibfnamefont {M.~O.}\ \bibnamefont
  {Ajeesh}}, \bibinfo {author} {\bibfnamefont {N.}~\bibnamefont {Kumar}},
  \bibinfo {author} {\bibfnamefont {F.}~\bibnamefont {Arnold}}, \bibinfo
  {author} {\bibfnamefont {C.}~\bibnamefont {Shekhar}}, \bibinfo {author}
  {\bibfnamefont {M.}~\bibnamefont {Naumann}}, \bibinfo {author} {\bibfnamefont
  {M.}~\bibnamefont {Schmidt}}, \bibinfo {author} {\bibfnamefont
  {M.}~\bibnamefont {Nicklas}},\ and\ \bibinfo {author} {\bibfnamefont
  {E.}~\bibnamefont {Hassinger}},\ }\bibfield  {title} {\bibinfo {title} {{On
  the search for the chiral anomaly in Weyl semimetals: the negative
  longitudinal magnetoresistance}},\ }\href
  {https://doi.org/10.1088/1367-2630/18/8/085006} {\bibfield  {journal}
  {\bibinfo  {journal} {New J. Phys.}\ }\textbf {\bibinfo {volume} {18}},\
  \bibinfo {pages} {085006} (\bibinfo {year} {2016})}\BibitemShut {NoStop}%
\bibitem [{\citenamefont {Naumann}\ \emph {et~al.}(2020)\citenamefont
  {Naumann}, \citenamefont {Arnold}, \citenamefont {Bachmann}, \citenamefont
  {Modic}, \citenamefont {Moll}, \citenamefont {S\"u\ss{}}, \citenamefont
  {Schmidt},\ and\ \citenamefont {Hassinger}}]{Naumann2020}%
  \BibitemOpen
  \bibfield  {author} {\bibinfo {author} {\bibfnamefont {M.}~\bibnamefont
  {Naumann}}, \bibinfo {author} {\bibfnamefont {F.}~\bibnamefont {Arnold}},
  \bibinfo {author} {\bibfnamefont {M.~D.}\ \bibnamefont {Bachmann}}, \bibinfo
  {author} {\bibfnamefont {K.~A.}\ \bibnamefont {Modic}}, \bibinfo {author}
  {\bibfnamefont {P.~J.~W.}\ \bibnamefont {Moll}}, \bibinfo {author}
  {\bibfnamefont {V.}~\bibnamefont {S\"u\ss{}}}, \bibinfo {author}
  {\bibfnamefont {M.}~\bibnamefont {Schmidt}},\ and\ \bibinfo {author}
  {\bibfnamefont {E.}~\bibnamefont {Hassinger}},\ }\bibfield  {title} {\bibinfo
  {title} {Orbital effect and weak localization in the longitudinal
  magnetoresistance of weyl semimetals nbp, nbas, tap, and taas},\ }\href
  {https://doi.org/10.1103/PhysRevMaterials.4.034201} {\bibfield  {journal}
  {\bibinfo  {journal} {Phys. Rev. Mater.}\ }\textbf {\bibinfo {volume} {4}},\
  \bibinfo {pages} {034201} (\bibinfo {year} {2020})}\BibitemShut {NoStop}%
\bibitem [{\citenamefont {Lv}\ \emph {et~al.}(2021)\citenamefont {Lv},
  \citenamefont {Qian},\ and\ \citenamefont {Ding}}]{Lv2021}%
  \BibitemOpen
  \bibfield  {author} {\bibinfo {author} {\bibfnamefont {B.~Q.}\ \bibnamefont
  {Lv}}, \bibinfo {author} {\bibfnamefont {T.}~\bibnamefont {Qian}},\ and\
  \bibinfo {author} {\bibfnamefont {H.}~\bibnamefont {Ding}},\ }\bibfield
  {title} {\bibinfo {title} {{Experimental perspective on three-dimensional
  topological semimetals}},\ }\href
  {https://doi.org/10.1103/RevModPhys.93.025002} {\bibfield  {journal}
  {\bibinfo  {journal} {Rev. Mod. Phys.}\ }\textbf {\bibinfo {volume} {93}},\
  \bibinfo {pages} {25002} (\bibinfo {year} {2021})}\BibitemShut {NoStop}%
\bibitem [{\citenamefont {Zhang}\ \emph {et~al.}(2023)\citenamefont {Zhang},
  \citenamefont {Jiang}, \citenamefont {Yun}, \citenamefont {Benally},
  \citenamefont {Peterson}, \citenamefont {Cresswell}, \citenamefont {Fan},
  \citenamefont {Lv}, \citenamefont {Yu}, \citenamefont {Barriocanal} \emph
  {et~al.}}]{Zhang2023}%
  \BibitemOpen
  \bibfield  {author} {\bibinfo {author} {\bibfnamefont {D.}~\bibnamefont
  {Zhang}}, \bibinfo {author} {\bibfnamefont {W.}~\bibnamefont {Jiang}},
  \bibinfo {author} {\bibfnamefont {H.}~\bibnamefont {Yun}}, \bibinfo {author}
  {\bibfnamefont {O.~J.}\ \bibnamefont {Benally}}, \bibinfo {author}
  {\bibfnamefont {T.}~\bibnamefont {Peterson}}, \bibinfo {author}
  {\bibfnamefont {Z.}~\bibnamefont {Cresswell}}, \bibinfo {author}
  {\bibfnamefont {Y.}~\bibnamefont {Fan}}, \bibinfo {author} {\bibfnamefont
  {Y.}~\bibnamefont {Lv}}, \bibinfo {author} {\bibfnamefont {G.}~\bibnamefont
  {Yu}}, \bibinfo {author} {\bibfnamefont {J.~G.}\ \bibnamefont {Barriocanal}},
  \emph {et~al.},\ }\bibfield  {title} {\bibinfo {title} {Robust negative
  longitudinal magnetoresistance and spin--orbit torque in sputtered pt3sn and
  pt3snxfe1-x topological semimetal},\ }\href@noop {} {\bibfield  {journal}
  {\bibinfo  {journal} {Nature communications}\ }\textbf {\bibinfo {volume}
  {14}},\ \bibinfo {pages} {4151} (\bibinfo {year} {2023})}\BibitemShut
  {NoStop}%
\bibitem [{\citenamefont {Chaou}\ \emph {et~al.}(2023)\citenamefont {Chaou},
  \citenamefont {Dwivedi},\ and\ \citenamefont {Breitkreiz}}]{Chaou2023}%
  \BibitemOpen
  \bibfield  {author} {\bibinfo {author} {\bibfnamefont {A.~Y.}\ \bibnamefont
  {Chaou}}, \bibinfo {author} {\bibfnamefont {V.}~\bibnamefont {Dwivedi}},\
  and\ \bibinfo {author} {\bibfnamefont {M.}~\bibnamefont {Breitkreiz}},\
  }\bibfield  {title} {\bibinfo {title} {Magnetic breakdown and chiral magnetic
  effect at {{Weyl-semimetal}} tunnel junctions},\ }\href
  {https://doi.org/10.1103/PhysRevB.107.L241109} {\bibfield  {journal}
  {\bibinfo  {journal} {Physical Review B}\ }\textbf {\bibinfo {volume}
  {107}},\ \bibinfo {pages} {L241109} (\bibinfo {year} {2023})}\BibitemShut
  {NoStop}%
\bibitem [{\citenamefont {Chaou}\ \emph {et~al.}(2024)\citenamefont {Chaou},
  \citenamefont {Dwivedi},\ and\ \citenamefont {Breitkreiz}}]{Chaou2023a}%
  \BibitemOpen
  \bibfield  {author} {\bibinfo {author} {\bibfnamefont {A.~Y.}\ \bibnamefont
  {Chaou}}, \bibinfo {author} {\bibfnamefont {V.}~\bibnamefont {Dwivedi}},\
  and\ \bibinfo {author} {\bibfnamefont {M.}~\bibnamefont {Breitkreiz}},\
  }\bibfield  {title} {\bibinfo {title} {Quantum oscillation signatures of
  interface fermi arcs},\ }\href {https://doi.org/10.1103/PhysRevB.110.035116}
  {\bibfield  {journal} {\bibinfo  {journal} {Phys. Rev. B}\ }\textbf {\bibinfo
  {volume} {110}},\ \bibinfo {pages} {035116} (\bibinfo {year}
  {2024})}\BibitemShut {NoStop}%
\bibitem [{\citenamefont {Dwivedi}(2018)}]{Dwivedi2018}%
  \BibitemOpen
  \bibfield  {author} {\bibinfo {author} {\bibfnamefont {V.}~\bibnamefont
  {Dwivedi}},\ }\bibfield  {title} {\bibinfo {title} {Fermi arc reconstruction
  at junctions between {{Weyl}} semimetals},\ }\href
  {https://doi.org/10.1103/PhysRevB.97.064201} {\bibfield  {journal} {\bibinfo
  {journal} {Physical Review B}\ }\textbf {\bibinfo {volume} {97}},\ \bibinfo
  {pages} {064201} (\bibinfo {year} {2018})}\BibitemShut {NoStop}%
\bibitem [{\citenamefont {Murthy}\ \emph {et~al.}(2020)\citenamefont {Murthy},
  \citenamefont {Fertig},\ and\ \citenamefont {Shimshoni}}]{Murthy2020}%
  \BibitemOpen
  \bibfield  {author} {\bibinfo {author} {\bibfnamefont {G.}~\bibnamefont
  {Murthy}}, \bibinfo {author} {\bibfnamefont {H.~A.}\ \bibnamefont {Fertig}},\
  and\ \bibinfo {author} {\bibfnamefont {E.}~\bibnamefont {Shimshoni}},\
  }\bibfield  {title} {\bibinfo {title} {Surface states and arcless angles in
  twisted {{Weyl}} semimetals},\ }\href
  {https://doi.org/10.1103/PhysRevResearch.2.013367} {\bibfield  {journal}
  {\bibinfo  {journal} {Phys. Rev. Res.}\ }\textbf {\bibinfo {volume} {2}},\
  \bibinfo {pages} {13367} (\bibinfo {year} {2020})}\BibitemShut {NoStop}%
\bibitem [{\citenamefont {Abdulla}\ \emph {et~al.}(2021)\citenamefont
  {Abdulla}, \citenamefont {Rao},\ and\ \citenamefont {Murthy}}]{Abdulla2021}%
  \BibitemOpen
  \bibfield  {author} {\bibinfo {author} {\bibfnamefont {F.}~\bibnamefont
  {Abdulla}}, \bibinfo {author} {\bibfnamefont {S.}~\bibnamefont {Rao}},\ and\
  \bibinfo {author} {\bibfnamefont {G.}~\bibnamefont {Murthy}},\ }\bibfield
  {title} {\bibinfo {title} {Fermi arc reconstruction at the interface of
  twisted {{Weyl}} semimetals},\ }\href
  {https://doi.org/10.1103/PhysRevB.103.235308} {\bibfield  {journal} {\bibinfo
   {journal} {Physical Review B}\ }\textbf {\bibinfo {volume} {103}},\ \bibinfo
  {pages} {235308} (\bibinfo {year} {2021})}\BibitemShut {NoStop}%
\bibitem [{\citenamefont {Kaushik}\ \emph {et~al.}()\citenamefont {Kaushik},
  \citenamefont {Robredo}, \citenamefont {Mathur}, \citenamefont {Schoop},
  \citenamefont {Jin}, \citenamefont {Vergniory},\ and\ \citenamefont
  {Cano}}]{Kaushik2022}%
  \BibitemOpen
  \bibfield  {author} {\bibinfo {author} {\bibfnamefont {S.}~\bibnamefont
  {Kaushik}}, \bibinfo {author} {\bibfnamefont {I.}~\bibnamefont {Robredo}},
  \bibinfo {author} {\bibfnamefont {N.}~\bibnamefont {Mathur}}, \bibinfo
  {author} {\bibfnamefont {L.~M.}\ \bibnamefont {Schoop}}, \bibinfo {author}
  {\bibfnamefont {S.}~\bibnamefont {Jin}}, \bibinfo {author} {\bibfnamefont
  {M.~G.}\ \bibnamefont {Vergniory}},\ and\ \bibinfo {author} {\bibfnamefont
  {J.}~\bibnamefont {Cano}},\ }\href {http://arxiv.org/abs/2207.14109}
  {\bibinfo {title} {Transport signatures of {{Fermi}} arcs at twin boundaries
  in {{Weyl}} materials}},\ \Eprint {https://arxiv.org/abs/2207.14109}
  {arxiv:2207.14109} \BibitemShut {NoStop}%
\bibitem [{\citenamefont {Buccheri}\ \emph {et~al.}(2022)\citenamefont
  {Buccheri}, \citenamefont {Egger},\ and\ \citenamefont
  {De~Martino}}]{Buccheri2022}%
  \BibitemOpen
  \bibfield  {author} {\bibinfo {author} {\bibfnamefont {F.}~\bibnamefont
  {Buccheri}}, \bibinfo {author} {\bibfnamefont {R.}~\bibnamefont {Egger}},\
  and\ \bibinfo {author} {\bibfnamefont {A.}~\bibnamefont {De~Martino}},\
  }\bibfield  {title} {\bibinfo {title} {Transport, refraction, and interface
  arcs in junctions of {{Weyl}} semimetals},\ }\href
  {https://doi.org/10.1103/PhysRevB.106.045413} {\bibfield  {journal} {\bibinfo
   {journal} {Physical Review B}\ }\textbf {\bibinfo {volume} {106}},\ \bibinfo
  {pages} {45413} (\bibinfo {year} {2022})}\BibitemShut {NoStop}%
\bibitem [{\citenamefont {Mathur}\ \emph {et~al.}(2023)\citenamefont {Mathur},
  \citenamefont {Yuan}, \citenamefont {Cheng}, \citenamefont {Kaushik},
  \citenamefont {Robredo},\ and\ \citenamefont {Maia}}]{Mathur2023}%
  \BibitemOpen
  \bibfield  {author} {\bibinfo {author} {\bibfnamefont {N.}~\bibnamefont
  {Mathur}}, \bibinfo {author} {\bibfnamefont {F.}~\bibnamefont {Yuan}},
  \bibinfo {author} {\bibfnamefont {G.}~\bibnamefont {Cheng}}, \bibinfo
  {author} {\bibfnamefont {S.}~\bibnamefont {Kaushik}}, \bibinfo {author}
  {\bibfnamefont {I.}~\bibnamefont {Robredo}},\ and\ \bibinfo {author}
  {\bibfnamefont {G.}~\bibnamefont {Maia}},\ }\bibfield  {title} {\bibinfo
  {title} {Atomically sharp internal interface in a chiral weyl semimetal
  nanowire},\ }\href {https://doi.org/10.1021/acs.nanolett.2c05100} {\bibfield
  {journal} {\bibinfo  {journal} {Nano Letters}\ }\textbf {\bibinfo {volume}
  {23}},\ \bibinfo {pages} {2695} (\bibinfo {year} {2023})}\BibitemShut
  {NoStop}%
\bibitem [{\citenamefont {Kundu}\ \emph {et~al.}(2023)\citenamefont {Kundu},
  \citenamefont {Fertig},\ and\ \citenamefont {Kundu}}]{Kundu2023}%
  \BibitemOpen
  \bibfield  {author} {\bibinfo {author} {\bibfnamefont {R.}~\bibnamefont
  {Kundu}}, \bibinfo {author} {\bibfnamefont {H.~A.}\ \bibnamefont {Fertig}},\
  and\ \bibinfo {author} {\bibfnamefont {A.}~\bibnamefont {Kundu}},\ }\bibfield
   {title} {\bibinfo {title} {Broken symmetry and competing orders in {{Weyl}}
  semimetal interfaces},\ }\href {https://doi.org/10.1103/PhysRevB.107.L041402}
  {\bibfield  {journal} {\bibinfo  {journal} {Physical Review B}\ }\textbf
  {\bibinfo {volume} {107}},\ \bibinfo {pages} {L041402} (\bibinfo {year}
  {2023})}\BibitemShut {NoStop}%
\bibitem [{\citenamefont {Breitkreiz}\ and\ \citenamefont
  {Brouwer}(2023)}]{Breitkreiz2023}%
  \BibitemOpen
  \bibfield  {author} {\bibinfo {author} {\bibfnamefont {M.}~\bibnamefont
  {Breitkreiz}}\ and\ \bibinfo {author} {\bibfnamefont {P.~W.}\ \bibnamefont
  {Brouwer}},\ }\bibfield  {title} {\bibinfo {title} {Fermi-arc metals},\
  }\href {https://doi.org/10.1103/PhysRevLett.130.196602} {\bibfield  {journal}
  {\bibinfo  {journal} {Phys. Rev. Lett}\ }\textbf {\bibinfo {volume} {130}},\
  \bibinfo {pages} {196602} (\bibinfo {year} {2023})}\BibitemShut {NoStop}%
\bibitem [{\citenamefont {Breitkreiz}\ and\ \citenamefont
  {Brouwer}(2019)}]{Breitkreiz2019}%
  \BibitemOpen
  \bibfield  {author} {\bibinfo {author} {\bibfnamefont {M.}~\bibnamefont
  {Breitkreiz}}\ and\ \bibinfo {author} {\bibfnamefont {P.~W.}\ \bibnamefont
  {Brouwer}},\ }\bibfield  {title} {\bibinfo {title} {Large contribution of
  {{Fermi}} arcs to the conductivity of topological metals},\ }\href
  {https://doi.org/10.1103/PhysRevLett.123.066804} {\bibfield  {journal}
  {\bibinfo  {journal} {Phys. Rev. Lett}\ }\textbf {\bibinfo {volume} {123}},\
  \bibinfo {pages} {066804} (\bibinfo {year} {2019})}\BibitemShut {NoStop}%
\bibitem [{\citenamefont {{Perez-Piskunow}}\ \emph {et~al.}(2021)\citenamefont
  {{Perez-Piskunow}}, \citenamefont {Bovenzi}, \citenamefont {Akhmerov},\ and\
  \citenamefont {Breitkreiz}}]{Piskunow2021}%
  \BibitemOpen
  \bibfield  {author} {\bibinfo {author} {\bibfnamefont {P.~M.}\ \bibnamefont
  {{Perez-Piskunow}}}, \bibinfo {author} {\bibfnamefont {N.}~\bibnamefont
  {Bovenzi}}, \bibinfo {author} {\bibfnamefont {A.~R.}\ \bibnamefont
  {Akhmerov}},\ and\ \bibinfo {author} {\bibfnamefont {M.}~\bibnamefont
  {Breitkreiz}},\ }\bibfield  {title} {\bibinfo {title} {Chiral anomaly trapped
  in weyl metals: {{Nonequilibrium}} valley polarization at zero magnetic
  field},\ }\href {https://doi.org/10.21468/SciPostPhys.11.2.046} {\bibfield
  {journal} {\bibinfo  {journal} {SciPost Phys.}\ }\textbf {\bibinfo {volume}
  {11}},\ \bibinfo {pages} {046} (\bibinfo {year} {2021})}\BibitemShut
  {NoStop}%
\bibitem [{\citenamefont {Lanzillo}\ \emph {et~al.}(2024)\citenamefont
  {Lanzillo}, \citenamefont {Bajpai},\ and\ \citenamefont
  {Chen}}]{Lanzillo2024}%
  \BibitemOpen
  \bibfield  {author} {\bibinfo {author} {\bibfnamefont {N.~A.}\ \bibnamefont
  {Lanzillo}}, \bibinfo {author} {\bibfnamefont {U.}~\bibnamefont {Bajpai}},\
  and\ \bibinfo {author} {\bibfnamefont {C.-T.}\ \bibnamefont {Chen}},\
  }\bibfield  {title} {\bibinfo {title} {{Topological semimetal interface
  resistivity scaling for vertical interconnect applications}},\ }\href
  {https://doi.org/10.1063/5.0200403} {\bibfield  {journal} {\bibinfo
  {journal} {Appl. Phys. Lett.}\ }\textbf {\bibinfo {volume} {124}},\ \bibinfo
  {pages} {181603} (\bibinfo {year} {2024})}\BibitemShut {NoStop}%
\bibitem [{\citenamefont {Leahy}\ \emph {et~al.}(2024)\citenamefont {Leahy},
  \citenamefont {Rice}, \citenamefont {Jiang}, \citenamefont {Paul},
  \citenamefont {Alberi},\ and\ \citenamefont {Nelson}}]{Leahy2024}%
  \BibitemOpen
  \bibfield  {author} {\bibinfo {author} {\bibfnamefont {I.~A.}\ \bibnamefont
  {Leahy}}, \bibinfo {author} {\bibfnamefont {A.~D.}\ \bibnamefont {Rice}},
  \bibinfo {author} {\bibfnamefont {C.-S.}\ \bibnamefont {Jiang}}, \bibinfo
  {author} {\bibfnamefont {G.}~\bibnamefont {Paul}}, \bibinfo {author}
  {\bibfnamefont {K.}~\bibnamefont {Alberi}},\ and\ \bibinfo {author}
  {\bibfnamefont {J.~N.}\ \bibnamefont {Nelson}},\ }\bibfield  {title}
  {\bibinfo {title} {Anisotropic weak antilocalization in thin films of the
  weyl semimetal taas},\ }\href {https://doi.org/10.1103/PhysRevB.110.054206}
  {\bibfield  {journal} {\bibinfo  {journal} {Phys. Rev. B}\ }\textbf {\bibinfo
  {volume} {110}},\ \bibinfo {pages} {054206} (\bibinfo {year}
  {2024})}\BibitemShut {NoStop}%
\bibitem [{\citenamefont {Khan}\ \emph {et~al.}(2025)\citenamefont {Khan},
  \citenamefont {Ramdas}, \citenamefont {Lindgren}, \citenamefont {Kim},
  \citenamefont {Won}, \citenamefont {Wu}, \citenamefont {Saraswat},
  \citenamefont {Chen}, \citenamefont {Suzuki}, \citenamefont {da~Jornada},
  \citenamefont {Oh},\ and\ \citenamefont {Pop}}]{Khan2025}%
  \BibitemOpen
  \bibfield  {author} {\bibinfo {author} {\bibfnamefont {A.~I.}\ \bibnamefont
  {Khan}}, \bibinfo {author} {\bibfnamefont {A.}~\bibnamefont {Ramdas}},
  \bibinfo {author} {\bibfnamefont {E.}~\bibnamefont {Lindgren}}, \bibinfo
  {author} {\bibfnamefont {H.-M.}\ \bibnamefont {Kim}}, \bibinfo {author}
  {\bibfnamefont {B.}~\bibnamefont {Won}}, \bibinfo {author} {\bibfnamefont
  {X.}~\bibnamefont {Wu}}, \bibinfo {author} {\bibfnamefont {K.}~\bibnamefont
  {Saraswat}}, \bibinfo {author} {\bibfnamefont {C.-T.}\ \bibnamefont {Chen}},
  \bibinfo {author} {\bibfnamefont {Y.}~\bibnamefont {Suzuki}}, \bibinfo
  {author} {\bibfnamefont {F.~H.}\ \bibnamefont {da~Jornada}}, \bibinfo
  {author} {\bibfnamefont {I.-K.}\ \bibnamefont {Oh}},\ and\ \bibinfo {author}
  {\bibfnamefont {E.}~\bibnamefont {Pop}},\ }\bibfield  {title} {\bibinfo
  {title} {Surface conduction and reduced electrical resistivity in ultrathin
  noncrystalline nbp semimetal},\ }\href
  {https://doi.org/10.1126/science.adq7096} {\bibfield  {journal} {\bibinfo
  {journal} {Science}\ }\textbf {\bibinfo {volume} {387}},\ \bibinfo {pages}
  {62} (\bibinfo {year} {2025})}\BibitemShut {NoStop}%
\bibitem [{\citenamefont {Kohn}\ and\ \citenamefont
  {Luttinger}(1957)}]{Kohn1957}%
  \BibitemOpen
  \bibfield  {author} {\bibinfo {author} {\bibfnamefont {W.}~\bibnamefont
  {Kohn}}\ and\ \bibinfo {author} {\bibfnamefont {J.}~\bibnamefont
  {Luttinger}},\ }\bibfield  {title} {\bibinfo {title} {{Quantum Theory of
  Electrical Transport Phenomena}},\ }\href
  {https://doi.org/10.1103/PhysRev.108.590} {\bibfield  {journal} {\bibinfo
  {journal} {Phys. Rev.}\ }\textbf {\bibinfo {volume} {108}},\ \bibinfo {pages}
  {590} (\bibinfo {year} {1957})}\BibitemShut {NoStop}%
\bibitem [{\citenamefont {Williams}\ \emph {et~al.}(2007)\citenamefont
  {Williams}, \citenamefont {DiCarlo},\ and\ \citenamefont
  {Marcus}}]{Williams2007}%
  \BibitemOpen
  \bibfield  {author} {\bibinfo {author} {\bibfnamefont {J.~R.}\ \bibnamefont
  {Williams}}, \bibinfo {author} {\bibfnamefont {L.}~\bibnamefont {DiCarlo}},\
  and\ \bibinfo {author} {\bibfnamefont {C.~M.}\ \bibnamefont {Marcus}},\
  }\bibfield  {title} {\bibinfo {title} {Quantum hall effect in a
  gate-controlled <i>p-n</i> junction of graphene},\ }\href
  {https://doi.org/10.1126/science.1144657} {\bibfield  {journal} {\bibinfo
  {journal} {Science}\ }\textbf {\bibinfo {volume} {317}},\ \bibinfo {pages}
  {638} (\bibinfo {year} {2007})}\BibitemShut {NoStop}%
\bibitem [{\citenamefont {Abanin}\ and\ \citenamefont
  {Levitov}(2007)}]{Abanin2007}%
  \BibitemOpen
  \bibfield  {author} {\bibinfo {author} {\bibfnamefont {D.~A.}\ \bibnamefont
  {Abanin}}\ and\ \bibinfo {author} {\bibfnamefont {L.~S.}\ \bibnamefont
  {Levitov}},\ }\bibfield  {title} {\bibinfo {title} {Quantized transport in
  graphene p-n junctions in a magnetic field},\ }\href
  {https://doi.org/10.1126/science.1144672} {\bibfield  {journal} {\bibinfo
  {journal} {Science}\ }\textbf {\bibinfo {volume} {317}},\ \bibinfo {pages}
  {641} (\bibinfo {year} {2007})}\BibitemShut {NoStop}%
\bibitem [{\citenamefont {Groth}\ \emph {et~al.}(2014)\citenamefont {Groth},
  \citenamefont {Wimmer}, \citenamefont {Akhmerov},\ and\ \citenamefont
  {Waintal}}]{Groth2014}%
  \BibitemOpen
  \bibfield  {author} {\bibinfo {author} {\bibfnamefont {C.~W.}\ \bibnamefont
  {Groth}}, \bibinfo {author} {\bibfnamefont {M.}~\bibnamefont {Wimmer}},
  \bibinfo {author} {\bibfnamefont {A.~R.}\ \bibnamefont {Akhmerov}},\ and\
  \bibinfo {author} {\bibfnamefont {X.}~\bibnamefont {Waintal}},\ }\bibfield
  {title} {\bibinfo {title} {Kwant: A software package for quantum transport},\
  }\href {https://doi.org/10.1088/1367-2630/16/6/063065} {\bibfield  {journal}
  {\bibinfo  {journal} {New Journal of Physics}\ }\textbf {\bibinfo {volume}
  {16}},\ \bibinfo {pages} {063065} (\bibinfo {year} {2014})}\BibitemShut
  {NoStop}%
\bibitem [{\citenamefont {Dwivedi}\ and\ \citenamefont
  {Chua}(2016)}]{Dwivedi2016a}%
  \BibitemOpen
  \bibfield  {author} {\bibinfo {author} {\bibfnamefont {V.}~\bibnamefont
  {Dwivedi}}\ and\ \bibinfo {author} {\bibfnamefont {V.}~\bibnamefont {Chua}},\
  }\bibfield  {title} {\bibinfo {title} {Of bulk and boundaries:
  {{Generalized}} transfer matrices for tight-binding models},\ }\href
  {https://doi.org/10.1103/PhysRevB.93.134304} {\bibfield  {journal} {\bibinfo
  {journal} {Physical Review B}\ }\textbf {\bibinfo {volume} {93}},\ \bibinfo
  {pages} {134304} (\bibinfo {year} {2016})}\BibitemShut {NoStop}%
\end{thebibliography}%

\appendix 

\section{Transfer matrices} 
\label{app:tmat}
\newcommand\zterm{\mu}

In this Appendix, we use transfer matrices to derive various quantities of interest for the model calculations based on the Landauer-Boltzmann approach, namely the group velocity of the interface Fermi arcs and their wavefunction overlap. Explicitly, we consider the Bloch Hamiltonians of the form
\begin{align} 
    \hlt_A &= \sin k_x \tau^x +  \eta_y^A(\ktrans) \tau^y + \left[ \zterm(\ktrans) - \cos k_x \right] \tau^z, 
\end{align} 
where $\zterm(\ktrans) = (1 + 2\cos k_0 - \cos k_y - \cos k_z)$ and $A \in \{\L,\R\}$. To construct the corresponding generalized transfer matrix along $x$, we first rewrite the Hamiltonian~as 
\begin{align} 
	\hlt_A(\vk_{3d}) = J \e^{-\i k_x} + M_A(\vk) + J^\dg \e^{\i k_x}, 
\end{align} 
where we identify the hopping and on-site matrices 
\[ 
    J = \frac12( \i\tau^x - \tau^z), \qquad 
    M_A = \eta_y^A(\ktrans) \tau^y + \zterm(\ktrans) \tau^z. 
\]
The real space Schr\"odinger equation takes the form of a recursion relation 
\begin{equation}
    (\ve - M_A) \psi_n = J \psi_{n+1} + J^\dg \psi_{n-1}. \label{eq:recur}
\end{equation}
As $J$ is singular, we follow the approach of \cite{Dwivedi2016a}, whereby we rewrite the recursion relation as  
\begin{equation}
    \psi_n = \green_A J \psi_{n+1} + \green_A J^\dg \psi_{n-1}. \label{eq:recur2}
\end{equation}
where $\green_A = (\ve - M_A)^{-1}$. Next, we compute the reduced singular value decomposition $J = \vv\cdot\vw^\dg$, whereby $\vv$ and $\vw$ form an orthonormal basis of $\cmplx^2$. Expanding the wavefunction as $\psi_n = \alpha_n \vv + \beta_n \vw$ with $\alpha_n, \beta_n \in\cmplx$ and taking inner products of Eq.~\eqref{eq:recur2} with $\vv$ and $\vw$, we get a pair of recursion relations for $\alpha_n$ and $\beta_n$, which can be cast in the transfer matrix form 
\begin{align}
     \Phi_{n+1} = \Tmat_A (\ve, \ktrans) \Phi_n, \qquad 
     \Phi_n = 
     \begin{pmatrix} 
        \beta_{n} \\ \alpha_{n-1}
     \end{pmatrix}, 
\end{align}
with 
\begin{equation} 
	\Tmat_A(\ve, \ktrans) = \frac1\zterm
	\begin{pmatrix}
		\ve^2 - \left(\eta_y^A\right)^2 - \zterm^2 \;\; & -\ve+\eta_y^A \\ 
		\ve+\eta_y^A        & -1 
	\end{pmatrix}. 
\label{TM}
\end{equation} 

To derive the interface modes, we briefly recap the approach of Ref. \cite{Dwivedi2018}. Assuming the interface lies at $n=0$ and the coupling across the interface is scaled by $\tu\in[0,1]$, the recursion relation of Eq.~\eqref{eq:recur} for $n=0,1$ becomes  
\begin{align}
     (\ve - M_\L) \psi_0 &= \tu J \psi_1 + J^\dg \psi_{-1}. \nonumber \\ 
     (\ve - M_\R) \psi_1 &= J \psi_2 + \tu J^\dg \psi_0. 
\end{align}
Again expanding $\psi_n$ in the $(\vv, \vw)$ basis, we can combine these equations into 
\begin{align}
    \Phi_1 &= \diag{\tu^{-1}, 1} \cdot \Tmat_\L \Phi_0,  \nonumber \\ 
    \Phi_2 &= \Tmat_\R \cdot \diag{1,\tu} \Phi_2, 
\end{align}
so that 
\begin{equation}
    \Phi_2 = \Tmat_\R \Kmat \Tmat_\L \Phi_0, \qquad 
    \Kmat = \diag{\tu^{-1}, \tu}.  
\end{equation}
For a state to be localized at the interface, it must decay exponentially on either side of the interface, so that $\Phi_0$ and $\Phi_2$ must be eigenvectors of $\Tmat_{\L/\R}$ with eigenvalues $\rho_{\L/\R}$, respectively. Using the relation between $\Phi_0$ and $\Phi_2$, we can rewrite 
\begin{align}
      \rho_\R \Phi_0 
      &= \left(\Tmat_\R \Kmat \Tmat_\L \right)^{-1} \Tmat_\R \left( \Tmat_\R \Kmat \Tmat_\L \right) \Phi_0  \nonumber \\ 
      &= \left( \Tmat_\L^{-1} \Kmat^{-1} \Tmat_\R \Kmat \Tmat_\L \right) \Phi_0,  
\end{align}
so that $\Phi_0$ is a simultaneous eigenvector of the matrices $\Tmat_\L$ and $\Tmat_\L^{-1} \Kmat^{-1} \Tmat_\R \Kmat \Tmat_\L$. A necessary condition for such a simultaneous eigenvector to exist is given by \cite{Dwivedi2018}
\begin{equation} 
	F(\ve, \ktrans) \equiv \det \left[ \Tmat_\L, \Kmat^{-1} \Tmat_\R \Kmat \right] = 0. 
	\label{matching_cond}
\end{equation} 
To derive the interface states, we need to solve this implicit relation between $\ve$ and $\ktrans$. We now describe the computation of various quantities of interest for the semiclassics. 

\subsection{Group velocity of the interface Fermi arcs}
To compute the group velocity of the interface Fermi arcs from the transfer matrices, we note that there exists an interface mode at momentum $\ktrans$ and energy $\ve$ if $F(\ve, \ktrans)=0$ (see Eq.~\eqref{matching_cond}). If $\ve+\delta\ve$ and $\ktrans+\delta\ktrans$ also satisfy this condition, then at linear order, we get 
\begin{equation} 
	\left[ \delta \ve \frac\partial{\partial\ve} + \delta\ktrans \cdot \nabla_{\ktrans} \right] F(\ve, \ktrans) = 0. 
\end{equation}
Thus, the group velocity is given by 
\begin{equation}
	\vv = \nabla_{\ktrans} \ve = - \frac{\nabla_{\ktrans} F(\ve, \ktrans)}{\partial_\ve  F(\ve, \ktrans)}.  
\end{equation} 
Having computed the Fermi arc for a given chemical potential, we can numerically evaluate this expression to obtain the velocity for each point on the Fermi arc. 

\subsection{Wavefunction overlap}
The transition probability for a disorder-induced scattering $\ktrans \to \ktrans'$ depends on the overlap within the interface layers of the interface modes with transverse momenta $\vk_{1,2}$. Explicitly, given the \emph{unnormalized} wavefunctions $\ket{\Psi}$ and $\ket{\Psi'}$ with $\ket{\Psi} \equiv \sum_n \psi_n \ket{n}$ etc, we need to compute 
\begin{equation}
    \braket{\Psi| \Psi'}_\text{int} = \frac{\psi_0^\dg \psi_0' + \psi_1^\dg \psi_1'}{\sqrt{\braket{\Psi|\Psi} \braket{\Psi'|\Psi'}}},  \label{overlap_def}
\end{equation}
where the sum is only over the interface layers $n=0,1$. 

\begin{figure}
\includegraphics[width=\columnwidth]{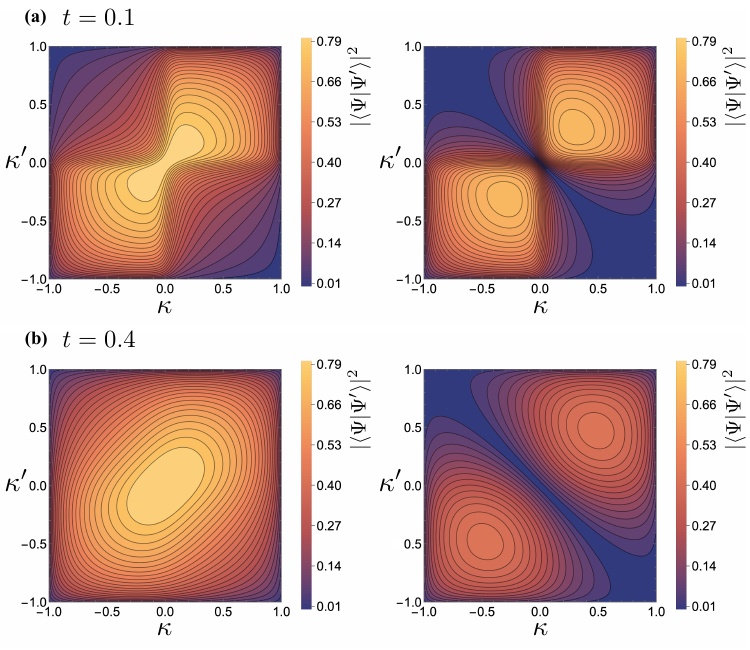}
    \caption{Overlap of Fermi-arc wavefunctions $\abs{\braket{\Psi| \Psi'}}^2$ for intra-arc (left) and inter-arc (right) scattering for $\varepsilon=0$ and tunneling parameter (a)  $t=0.1$ and (b) $t=0.4$. }
    \label{fig:overlap}
\end{figure}

We now describe the explicit computation of these states from the eigenvalues and eigenvectors of the transfer matrix. Note that an interface states with energy $\ve$ exists for a given $\ktrans$ only if Eq.~\eqref{matching_cond} holds, \ie, only if $\ker{\left[ \Tmat_\L, K^{-1} \Tmat_\R K \right]}$ is nontrivial. Let ${\left[ \Tmat_\L, \Kmat^{-1} \Tmat_\R \Kmat \right]} \varphi = 0$ for a $\varphi \in \cmplx^2$, so that $\varphi$ satisfies 
\begin{equation}
    \Tmat_\L \varphi = \rho_\L \varphi \qquad 
    \Kmat^{-1} \Tmat_\R \Kmat \varphi = \rho_\R \varphi.  
\end{equation}
These can be used to compute 
\begin{equation}
    \Phi_n = 
    \begin{cases}
        \rho_\L^n \varphi, & n \leq 0, \\ 
        \rho_\L \, \diag{\tu^{-1}, 1} \varphi, & n=1, \\ 
        \rho_\L \rho_\R^{n-1} \Kmat\varphi, & n \geq 2. 
    \end{cases}
\end{equation}
Similarly, let $\varphi'$ be the corresponding vector for $\ktrans'$, from which we can compute $\Phi_n'$. 
We next seek to rewrite Eq.~\eqref{overlap_def} in terms of $\Phi_n$. We first compute the numerator as 
\begin{align}
    \sum_{n=0}^1 \psi_n^\ast \psi_n' 
    &= \sum_{n=0}^1  \left( \alpha_n^\ast \vv^\dg + \beta_n^\ast \vw^\dg \right) \left( \alpha_n' \vv + \beta_n' \vw \right) \nonumber \\
    &= \sum_{n=0}^1  \alpha_n^\ast \alpha_n' + \beta_n^\ast \beta_n',   
\end{align}
where we have used the orthonormality of $\vv$ and $\vw$. Setting $\varphi = (\varphi_1, \varphi_2)^T$, we identify $\alpha_0 = \rho_\L \varphi_2$, $\beta_0 = \varphi_1$, $\alpha_1 = \rho_\L \rho_\R \tu \varphi_2$, and $\beta_1 = \rho_\L \tu^{-1} \varphi_1$ and similarly for $\varphi'$. Thus, 
\begin{align}
    \sum_{n=0}^1 \psi_n^\ast \psi_n' 
    &= \left( 1 + \tu^{-2} \rho_\L^\ast \rho_\L' \right) \varphi_1^\ast \varphi_1' \nonumber \\ 
    & \quad + \rho_\L^\ast \rho_\L' \left( 1 + \tu^2 \rho_\R^\ast \rho_\R' \right) \varphi_2^\ast \varphi_2'. 
    \label{overlap_phi}
\end{align}
Similarly, we compute the normalization constant as 
\begin{align}
    \braket{\Psi|\Psi} 
    &= \sum_{n=-\infty}^\infty \abs{\Psi_n}^2
    = \sum_{n=-\infty}^\infty \left( \abs{\alpha_n}^2 + \abs{\beta_n}^2 \right) \nonumber \\ 
    &= \sum_{n=-\infty}^\infty \abs{\Phi_n}^2 \equiv \varphi^\dg \mathcal{O} \varphi, 
    \label{norm_phi}
\end{align}
where we have set 
\begin{align}
    \mathcal{O} 
    &= \sum_{n=-\infty}^0 \abs{\rho_\L}^{2n} + \abs{\rho_\L}^2 \diag{\tu^{-1}, 1}^2  \nonumber \\ 
    & \qquad + \abs{\rho_\L}^2 \Kmat^2 \sum_{n=2}^\infty \abs{\rho_\R}^{2(n-1)} \nonumber \\ 
    &= \abs{\rho_\L}^2 \text{diag}\left\{ \frac1{\abs{\rho_\L}^2 - 1} + \frac{\tu^{-2}}{1-\abs{\rho_\R}^2},  \right. \nonumber \\ 
    & \qquad \left. \frac{\abs{\rho_\L}^2}{\abs{\rho_\L}^2 - 1} + \frac{\tu^2 \abs{\rho_\R}^2}{1-\abs{\rho_\R}^2}  \right\}. 
\end{align}
Note that this matrix is positive-definite, since for a localized mode, we must have $\abs{\rho_{\L}}>1$ and $\abs{\rho_{\R}}<1$. Substituting Eqns.~\eqref{overlap_phi} and \eqref{norm_phi} in \eqref{overlap_def} then yields the desired overlaps. The result is shown in Fig.\ \ref{fig:overlap}

\end{document}